\documentclass{SciPost}

\IfFileExists{mathdesign.sty}{\usepackage[bitstream-charter]{mathdesign}}{\usepackage{lmodern}\usepackage{amssymb}}
\DeclareSymbolFont{usualmathcal}{OMS}{cmsy}{m}{n}
\DeclareSymbolFontAlphabet{\mathcal}{usualmathcal}

\usepackage{amsmath}
\usepackage{amsthm}

\usepackage{graphicx}
\usepackage{dcolumn}
\usepackage{bm}
\IfFileExists{dsfont.sty}{\usepackage{dsfont}}{}
\usepackage{slashed}
\usepackage{mathrsfs}
\IfFileExists{bbm.sty}{\usepackage{bbm}}{\let\mathbbm\mathbf}

\usepackage[all]{xy}
\usepackage{tikz}

\IfFileExists{physics.sty}{\usepackage{physics}}{}
\usepackage{enumitem}
\IfFileExists{soul.sty}{\usepackage{soul}}{}
\usepackage{float}
\usepackage{booktabs}

\usepackage{hyperref}
\hypersetup{
    colorlinks=true,
    linkcolor={red!50!black},
    citecolor={blue!50!black},
    urlcolor={blue!80!black}
}

\fancypagestyle{SPstyle}{
    \fancyhf{}
    \lhead{
        \colorbox{scipostblue}{
            \bfseries\color{white} SciPost Physics 
        }
    }
    \rhead{
        \bfseries\color{scipostdeepblue} Submission 
    }
    
    \fancyfoot[C]{\textbf{\thepage}}
}

\newcommand{\be}{\begin{equation}}
\newcommand{\ee}{\end{equation}}

\newcommand{\bea}{\begin{align}}
\newcommand{\eea}{\end{align}}

\def\inbar{\,\vrule height1.5ex width.4pt depth0pt}
\def\IR{\relax{\rm I\kern-.18em R}}
\def\IC{\relax\hbox{$\inbar\kern-.3em{\rm C}$}}

\begin{document}

\pagestyle{SPstyle}

\begin{center}{\Large \textbf{\color{scipostdeepblue}{
Toward a Holographic Representation Theory\\of\\the Bosonic de Sitter Discrete Series
}}}\end{center}

\begin{center}\textbf{
    Maryam Bajalan\textsuperscript{1$\star$},
    Jean-Pierre Gazeau\textsuperscript{2,3$\dagger$} and
    Hamed Pejhan\textsuperscript{1$\ddagger$}
}\end{center}

\begin{center}
    {\bf 1} Institute of Mathematics and Informatics, Bulgarian Academy of Sciences, Acad. G. Bonchev Str. Bl. 8, 1113, Sofia, Bulgaria \\
    {\bf 2} Universit\'e Paris Cit\'e, CNRS, Astroparticule et Cosmologie, F-75013 Paris, France \\
    {\bf 3} Faculty of Mathematics, University of Bia\l ystok, 15-245 Bia\l ystok, Poland
    \\[\baselineskip]
    $\star$ \href{mailto:email1}{\small maryam.bajalan@math.bas.bg}\,,\quad
    $\dagger$ \href{mailto:email2}{\small gazeau@apc.in2p3.fr}\,,\quad
    $\ddagger$ \href{mailto:email2}{\small pejhan@math.bas.bg}
\end{center}

\textbf{\boldmath{
\begin{abstract}
    We propose a program toward a holographic representation theory of the non-scalar bosonic discrete series $\Pi^\pm_{p,q}$, $p,q\in\mathbb{N}$, $1\leqslant q\leqslant p$, of the de Sitter ($\mathrm{dS}_{4}$) group $G=\mathrm{SO}_0(1,4)$. The proposal exploits two complementary realizations of the same unitary irreducible representation (UIR): a bounded-domain realization on an open ball $\mathbb{B}$ modeling the hyperbolic space $G/\mathrm{SO}(4)$, with $\partial\mathbb{B}\simeq\mathbb{S}^{3}$, and a covariant spacetime realization on $\mathrm{dS}_{4}$, with conformal boundaries $\mathscr{I}^{\pm}\simeq\mathbb{S}^{3}$, in which the physical UIR is obtained through the corresponding gauge reduction. This leads to two boundary constructions and an explicit compatibility problem, 
    \begin{align*}
        \mathbb{B}
        \;\overset{?\mathrm{hol}}{\longleftrightarrow}\;
        \partial\mathbb{B}\simeq\mathbb{S}^{3}
        \quad\overset{?}{\simeq}\quad
        \mathscr{I}^{\pm}\simeq\mathbb{S}^{3}
        \;\overset{?\mathrm{hol}}{\longleftrightarrow}\;
        \mathrm{dS}_{4} \,.
    \end{align*}
    The two outer problems seek explicit intertwiners between the bulk and boundary realizations, with their invariant norms, constraints, and physical quotients properly identified. The middle problem asks how to construct and normalize a unitary intertwiner between the two boundary realizations once both have been shown to carry the same UIR. The bounded picture exposes the unitary and harmonic structure of the representation, while the spacetime picture makes its covariant and gauge-theoretic structure manifest, allowing the two constructions to constrain and illuminate one another. The established scalar family $\Pi_{p,0}$, $p\in\mathbb{N}$, with its spherical realization on $\mathbb{S}^{3}$, provides the guiding prototype. Particular attention is given to the strictly massless vector representation $\Pi^\pm_{1,1}$, the strictly massless spin-$2$ (graviton) representation $\Pi^\pm_{2,2}$, and the partially massless spin-$2$ representation $\Pi^\pm_{2,1}$. Establishing the proposed correspondences for these representative sectors would provide concrete foundations for a common holographic representation theory of the bosonic discrete series and advance the understanding of holography for gravitational degrees of freedom in $\mathrm{dS}_{4}$ spacetime.
\end{abstract}
}}

\vspace{\baselineskip}

\noindent\textcolor{white!90!black}{%
\fbox{\parbox{0.975\linewidth}{%
\textcolor{white!40!black}{\begin{tabular}{lr}%
  \begin{minipage}{0.55\textwidth}%
    {\small Copyright attribution to authors. \newline
    This work is a submission to SciPost Physics. \newline
    License information to appear upon publication. \newline
    Publication information to appear upon publication.}
  \end{minipage} & \begin{minipage}{0.35\textwidth}
    {\small Received Date \newline Accepted Date \newline Published Date}%
  \end{minipage}
\end{tabular}}
}}}

%\linenumbers

%%%%%%%%%%%%%%%%%%%%%%%%%%%%%%%%%%%%%%%%%%%%%%%%%%%%%
%%%%%%%%%%%%%%%%%%%%%%%%%%%%%%%%%%%%%%%%%%%%%%%%%%%%%
%%%%%%%%%%%%%%%%%%%%%%%%%%%%%%%%%%%%%%%%%%%%%%%%%%%%%
\section{Introduction}
\label{sec:intro}

The discrete series of the de Sitter ($\mathrm{dS}_{4}$) group $\mathrm{SO}_0(1,4)$ occupies an exceptional position within its unitary dual. Its bosonic UIRs encompass the strictly massless, partially massless, and exceptional scalar representations. All these representations, including the scalar family, admit covariant field realizations with gauge invariance. In the realizations considered here, the gauge structure is encoded in indecomposable representations and Gupta--Bleuler constructions, with the physical UIR carried by the corresponding gauge quotient \cite{Dixmier, Takahashi, Gazeau2000, takVec, GravGB, GravPartially, Gazeau2024, Bajalan2026, bajalan2026araki, Gazeau2026, Gazeau2025}. Conformal symmetry provides further structure for the strictly massless and certain partially massless fields; the exceptional scalar family considered here is distinct from the conformally coupled scalar field. In discussing this additional symmetry, one must distinguish the extension of a physical $\mathrm{dS}_{4}$ UIR to a unitary representation of the conformal group from the conformal invariance of its field equations \cite{Barut, Todorov, Mack, Gazeau2026, Gazeau2025, DeserWaldron2004}. Relating these statements requires determining whether conformal transformations preserve the space of physical solutions and the gauge-equivalence relation, and whether the resulting action on the physical quotient preserves its inner product.

Concrete conformal field constructions provide further precedents. Huguet, Queva and Renaud treated conformally coupled scalar fields on $\mathrm{dS}_{4}$, anti-$\mathrm{dS}_{4}$, and Minkowski spaces as part of a construction valid in arbitrary spacetime dimension \cite{HuguetQuevaRenaud2006}, and exhibited a Weyl-induced intertwiner between the Minkowski and $\mathrm{dS}_{4}$ realizations of the conformal scalar representation \cite{HuguetQuevaRenaud2008}. This conformally coupled scalar representation is distinct from the exceptional scalar family used as our holographic prototype. For the photon, Faci, Huguet, Queva and Renaud constructed a conformally covariant Gupta--Bleuler quantization in $\mathrm{dS}_{4}$ spacetime \cite{FaciHuguetQuevaRenaud2009}. These works supply concrete precedents for conformal covariance and equivalence of field realizations; the particular boundary maps proposed here remain to be constructed, and their compatibility remains to be established.

The proposed program builds on an established literature on bulk--boundary relations and $\mathrm{dS}_{4}$ representation theory. Strominger's dS/CFT proposal and the higher-spin realization proposed by Anninos, Hartman and Strominger provide the broader holographic setting \cite{Strominger2001, Anninos2017}. Invariant kernels and intertwining operators supply the representation-theoretic foundations \cite{KnappStein1971, Dobrev1977}, while Dobrev explicitly formulated bulk--boundary equivalence through intertwiners in the Euclidean AdS setting \cite{Dobrev1999}. More recent treatments of UIRs relevant to $\mathrm{dS}_{4}$, including exceptional and discrete-series representations, are provided by Sun and Schaub \cite{Sun2025, Schaub2024}. For massive scalar fields, \c{S}eng\"or and Skordis investigated late-time boundary unitarity and the associated invariant norms \cite{SengorSkordis2020}. Building on these precedents, our objective is to develop a holographic representation-theoretic framework for the non-scalar bosonic discrete series, with particular attention to the invariant Hilbert structures and the gauge reductions through which the physical UIRs emerge.

From a holographic perspective, the gauge structure of these representations poses a central question: how can bulk and boundary realizations be related while preserving the same physical UIR? In the covariant gauge-field realizations considered here, the physical UIR is carried by a quotient of a larger indecomposable representation, obtained through Gupta--Bleuler reduction. Extracting asymptotic data is therefore only part of the problem. The boundary construction must also respect gauge equivalence, identify the induced $\mathrm{SO}_0(1,4)$ action, and recover the positive-definite invariant inner product on the physical quotient. An explicit unitary intertwiner must then relate the resulting physical boundary realization to its bulk counterpart. These requirements are particularly significant for the non-scalar bosonic discrete series, which includes the strictly massless and partially massless representations associated with the graviton and the partially massless graviton. The interplay between gauge reduction and unitary bulk--boundary equivalence motivates the framework developed below.

To address this problem, we propose to exploit two complementary realizations of the same physical $\mathrm{dS}_{4}$ UIR. The first is a bounded realization on an open ball $\mathbb{B}$ modeling the hyperbolic space $\mathrm{SO}_0(1,4)/\mathrm{SO}(4)$, whose boundary is the three-sphere \cite{Takahashi,Gazeau2024},
\begin{align}\label{eq:compact-boundary}
    \mathbb{B} \longrightarrow \partial\mathbb{B} \simeq \mathbb{S}^{3} \,.
\end{align}
The second is the covariant spacetime realization on $\mathrm{dS}_{4}$, in which the physical UIR emerges through the gauge reduction described above. Its future and past conformal boundaries are likewise three-spheres,
\begin{align}\label{eq:ds-boundary}
    \mathrm{dS}_{4} \longrightarrow \mathscr{I}^{\pm} \simeq \mathbb{S}^{3} \,.
\end{align}
These geometrically distinct settings suggest two complementary boundary constructions. Starting from the bounded realization, one seeks an explicit correspondence with a boundary realization on $\partial\mathbb{B}$. Starting from the covariant spacetime realization, one seeks a correspondence with asymptotic data on $\mathscr{I}^{\pm}$ that descends consistently to the physical gauge quotient. Existing boundary subquotient realizations and Szeg\"o-type intertwining constructions provide a starting point on the symmetric-space side \cite{KnappWallach1976, KnappWallach1980, Seppanen2009}. The task is to identify and normalize the maps appropriate to the representations considered here, together with their boundary carrier spaces, induced group actions, and invariant Hilbert structures. The common boundary geometry $\mathbb{S}^{3}$ then makes it possible to formulate the next question: how can the two resulting boundary realizations be identified explicitly?

Once the two boundary constructions have been established independently, the remaining task is to construct an explicit unitary intertwiner between them, identifying their invariant inner products and the physical quotient structures through which they carry the same UIR. Their common boundary geometry $\mathbb{S}^{3}$ does not by itself establish representation-theoretic equivalence. Although abstract unitary equivalence follows once both realizations are shown to carry the same UIR, its explicit implementation must still be determined. Constructing the two holographic correspondences and identifying their boundary realizations in this precise sense constitutes the central objective of the proposed program.

The strength of this parallel approach lies in the complementary information supplied by the two realizations. The covariant spacetime picture makes $\mathrm{dS}_{4}$ covariance, field equations, gauge structure, and asymptotic behavior explicit, whereas the bounded picture provides direct access to the unitary structure, invariant inner product, and harmonic decomposition of the physical UIR. Their respective boundary constructions would allow information obtained in one picture to guide and constrain the other, particularly in identifying the physical boundary carrier, determining its invariant norm, and implementing the Gupta--Bleuler reduction. An explicit unitary intertwiner between the resulting boundary realizations would then relate harmonic data from the ball to asymptotic spacetime data, potentially through an intrinsic boundary integral kernel. The common $\mathbb{S}^{3}$ would thereby provide the setting in which these complementary structures are explicitly matched.

The scalar discrete-series family provides a concrete prototype for this program \cite{Bajalan2026}. In this case, the counterpart of the non-scalar bounded realization is defined directly on $\mathbb{S}^{3}$, rather than on the interior of $\mathbb{B}$; we refer to it as the spherical realization. The corresponding $\mathrm{dS}_{4}$-covariant spacetime realization admits an explicit holographic correspondence with the future and past conformal boundaries $\mathscr{I}^{\pm}\simeq\mathbb{S}^{3}$. Upon passing to the physical quotients, an explicit unitary intertwiner identifies the boundary representation obtained from the spacetime construction with the independently defined spherical realization of the same scalar UIR. This correspondence matches the group actions and invariant Hilbert structures while accounting for the gauge sectors factored out in forming the physical quotients. The scalar case thus demonstrates how the physical representation carried by a covariant spacetime realization can be recovered on a boundary carrier.

For the non-scalar bosonic discrete series, an additional step is needed because the bounded realization is defined inside $\mathbb{B}$ rather than directly on $\mathbb{S}^{3}$. One must determine which boundary data on $\partial\mathbb{B}\simeq\mathbb{S}^{3}$ carry the same UIR and construct an explicit unitary map relating them to the ball realization. Existing Szeg\"o-type transforms, which map suitable boundary data to representations realized in the interior, provide a starting point \cite{KnappWallach1976, KnappWallach1980, Seppanen2009}. To use them here, one must specify the allowed boundary data, impose the required constraints, factor out any null directions, and identify the invariant inner product on the resulting space. The transform must then be normalized to preserve this inner product, and its inverse must recover the boundary data from the ball realization. Completing these steps would establish the ball--boundary correspondence. Its comparison with the spacetime construction requires a second boundary map, compatible with the gauge structure of the covariant fields.

For this spacetime-side problem, existing boundary descriptions of massless and partially massless fields provide useful precedents. Dolan, Nappi and Witten related partially massless AdS fields to constrained conformal boundary operators \cite{DolanNappiWitten2001}, while Bekaert and Grigoriev developed an ambient, gauge-covariant treatment of boundary values for massless and partially massless AdS fields \cite{BekaertGrigoriev2013}. These constructions offer methods for handling boundary constraints and gauge structures, although their adaptation to $\mathrm{dS}_{4}$ requires careful treatment of the different real forms and unitarity properties involved. In the present setting, the spacetime boundary map must descend consistently to the physical Gupta--Bleuler quotient and transport its positive-definite invariant inner product to the boundary. The resulting realization can then be explicitly intertwined with the one obtained from the ball.

The present work develops the representation-theoretic groundwork for this program, leaving the complete non-scalar boundary constructions for subsequent investigation. We focus on three representative UIRs: the strictly massless vector $\Pi^\pm_{1,1}$, the graviton $\Pi^\pm_{2,2}$, and the partially massless graviton $\Pi^\pm_{2,1}$. For each, we examine the bounded and $\mathrm{dS}_{4}$-covariant spacetime realizations in parallel, specifying their carrier spaces, invariant inner products, and group actions, together with the indecomposable structures and Gupta--Bleuler reductions of the spacetime picture. Guided by the established scalar prototype, this comparison identifies the data and compatibility conditions needed for the proposed boundary maps.

The program can be summarized by three interconnected problems:
\begin{align} \label{eq:program}
    \boxed{
        \mathbb{B} \;\overset{?\mathrm{hol}}{\longleftrightarrow}\; \partial\mathbb{B} \simeq \mathbb{S}^{3}
        \quad\overset{?}{\simeq}\quad
        \mathscr{I}^{\pm} \simeq \mathbb{S}^{3}
        \;\overset{?\mathrm{hol}}{\longleftrightarrow}\; \mathrm{dS}_{4}
        }
\end{align}
The outer question marks denote the two bulk--boundary constructions, each required to intertwine the $\mathrm{SO}_0(1,4)$ actions and preserve the physical invariant Hilbert structure, with the spacetime map descending consistently through gauge reduction. Once both boundary realizations have been shown to carry the same UIR, their abstract unitary equivalence follows. The middle question mark therefore asks for its explicit implementation: a normalized unitary intertwiner relating the two independently obtained boundary realizations.

A successful implementation of this program would clarify which aspects of the twofold boundary correspondence extend uniformly across the non-scalar bosonic discrete series and which depend on spin and gauge symmetry. The vector, graviton, and partially massless graviton sectors would provide concrete tests of this common structure. By explicitly relating their physical bulk and boundary realizations, the construction in Eq. \eqref{eq:program} could establish a representation-theoretic foundation for exploring holography for gravitational degrees of freedom and, ultimately, for developing a holographic formulation of quantum field theory in $\mathrm{dS}_{4}$ spacetime.

%%%%%%%%%%%%%%%%%%%%%%%%%%%%%%%%%%%%%%%%%%%%%%%%%%%%%
%%%%%%%%%%%%%%%%%%%%%%%%%%%%%%%%%%%%%%%%%%%%%%%%%%%%%
%%%%%%%%%%%%%%%%%%%%%%%%%%%%%%%%%%%%%%%%%%%%%%%%%%%%%
\section{Discrete-Series UIRs: Carrier Geometry, Orbits and Spacetime Realizations} \label{sec:discrete-realizations}

The program formulated in Eq. \eqref{eq:program} relates the bounded and covariant spacetime realizations of the same physical $\mathrm{dS}_{4}$ UIR. To prepare for their boundary constructions, we first specify the representations under consideration and distinguish the geometries underlying their realizations, including the role of coadjoint orbits. We use the Dixmier labels $(p,q)$ \cite{Dixmier} and focus on the bosonic discrete series. Half-integer-spin representations, when included in the general classification, are understood as representations of the double cover $\mathrm{Spin}_0(1,4)$.

The two independent Casimir operators provide invariant labels for the UIRs:
\begin{align} \label{eq:discrete-casimirs-def}
    Q^{(1)} = -\frac{1}{2}L_{ab}L^{ab}
    \,, \qquad
    Q^{(2)} = -W_aW^a
    \,, \qquad
    W_a = -\frac{1}{8}\epsilon_{abcde}L^{bc}L^{de} \,.
\end{align}
Here $a,b,\cdots=0,1,2,3,4$, indices are raised and lowered with $\eta_{ab}=\operatorname{diag}(1,-1,-1,-1,-1)$, and $\epsilon_{abcde}$ is the totally antisymmetric symbol normalized by $\epsilon_{01234}=1$. We use Hermitian generators $L_{ab}=-L_{ba}$, with commutation relations 
\begin{align} \label{eq:so14-commutation}
    [L_{ab},L_{cd}]
    =
    \mathrm{i}\left(
        \eta_{ac}L_{bd}
        +\eta_{bd}L_{ac}
        -\eta_{ad}L_{bc}
        -\eta_{bc}L_{ad}
    \right) \,.
\end{align}

In the Dixmier notation, the Casimir eigenvalues are
\begin{align} \label{eq:discrete-casimirs}
    \langle Q^{(1)}\rangle = -p(p+1)-(q+1)(q-2)
    \,, \qquad
    \langle Q^{(2)}\rangle = -p(p+1)q(q-1) \,.
\end{align}
For the non-scalar discrete-series UIRs $\Pi^\pm_{p,q}$, the labels satisfy
\begin{align}\label{range}
    p,q\in\frac{1}{2}\mathbb{N}
    \,, \qquad
    1\leqslant q\leqslant p
    \,, \qquad
    p-q\in\mathbb{N}_0 \,.
\end{align}
The bosonic representations considered here have integer $p,q$. The scalar family $\Pi_{p,0}$ instead has $p\in\mathbb{N}$ and $q=0$, with Casimir eigenvalues given by the same formulas:
\begin{align}
    \langle Q^{(1)}\rangle = 2-p(p+1)
    \,, \qquad
    \langle Q^{(2)}\rangle = 0 \,.
\end{align}
We retain the conventional designation ``scalar discrete series'' for this exceptional family, while distinguishing it from the square-integrable discrete series discussed below.

Within the non-scalar bosonic discrete series, the representations $\Pi^\pm_{p,p}$, $p\in\mathbb{N}$, form the strictly massless family, with spin $s=p$ and the two signs distinguishing the helicity sectors \cite{Gazeau2026, Gazeau2024, Gazeau2025, Barut, Todorov, Mack}. In particular, $\Pi^\pm_{1,1}$ corresponds to the photon \cite{takVec}, whereas $\Pi^\pm_{2,2}$ corresponds to the graviton \cite{GravGB}. The representations $\Pi^\pm_{p,q}$ with $1\leqslant q<p$ form the partially massless family \cite{Gazeau2024}. Higuchi and Deser and Waldron established the field-theoretic foundations of the spin-$2$ unitarity bound and partially massless gauge invariance \cite{Higuchi1987, DeserWaldron2001}. Conformal invariance of the maximal-depth partially massless family in four dimensions was studied in Ref. \cite{DeserWaldron2004}. The simplest bosonic partially massless representation is $\Pi^\pm_{2,1}$, associated with the partially massless spin-$2$ field \cite{GravPartially}.

The non-scalar representations in Eq. \eqref{range} belong to the genuine, square-integrable discrete series. The limiting representations with $q=\frac{1}{2}$, which occur on the double cover, are not square-integrable and are excluded from this range.

The scalar family $\Pi_{p,0}$, $p\in\mathbb{N}$, requires a separate qualification. Although traditionally included in the discrete series in the notation adopted here, these representations are not square-integrable; we refer to them more precisely as exceptional scalar subquotients. A singular infinitesimal character alone does not justify identifying them as limits of discrete series. Their carrier geometry also differs from that of the non-scalar bounded realization: as outlined in the Introduction, the scalar family is realized directly on $\mathbb{S}^{3}$, whereas the non-scalar ball model is defined on the interior of $\mathbb{B}$. The scalar spherical realization will therefore be treated separately.

For the non-scalar representations, it is useful to distinguish four geometric settings: (i) the bounded ball model of the Riemannian symmetric space $G/K$, with $G=\mathrm{SO}_0(1,4)$ and $K=\mathrm{SO}(4)$; (ii) unbounded models of the same symmetric space, related to the ball model by coordinate changes and the corresponding bundle transformations; (iii) coadjoint orbits, viewed as symplectic phase spaces with stabilizers determined by the chosen coadjoint elements; and (iv) covariant field realizations on Lorentzian $\mathrm{dS}_{4}$ spacetime. The first two describe the same base manifold in different geometric models. Coadjoint orbits and spacetime have different roles and must be distinguished from this symmetric-space carrier. In particular, passing from the ball model to a covariant spacetime realization is not a coordinate transformation: relating the two requires an intertwiner between their realizations of the same UIR, with the spacetime UIR identified through the appropriate physical gauge quotient.

%%%%%%%%%%%%%%%%%%%%%%%%%%%%%%%%%%%%%%%%%%%%%%%%%%%%%
%%%%%%%%%%%%%%%%%%%%%%%%%%%%%%%%%%%%%%%%%%%%%%%%%%%%%
\subsection{Bounded and Unbounded Models of the Symmetric-Space Carrier} \label{subsec:compact-realization}

Let $G=\mathrm{SO}_0(1,4)$ and $K=\mathrm{SO}(4)$ be its maximal compact subgroup. The quotient
\begin{align}\label{eq:carrier-geometry}
    X=G/K\simeq\mathbb H^4\simeq\mathbb B
    \,, \qquad
    \mathbb B=\{z\in\mathbb R^4 \;\colon\;\; |z|<1\}
\end{align}
is four-dimensional real hyperbolic space, with $\mathbb B$ understood as its Poincar\'e ball model. Passing to the double cover of $G$ and the corresponding preimage of $K$ leaves this symmetric space unchanged.

The ball $\mathbb B$ is bounded but non-compact. Its Euclidean closure provides a compactification whose boundary represents the ideal boundary of hyperbolic space:
\begin{align}\label{eq:compact-ball-boundary}
    \partial\mathbb B\simeq\mathbb S^3 \,.
\end{align}
We therefore use ``bounded realization'' for the ball model. The term ``compact'' is reserved for genuinely compact spaces, such as $K$ and $\mathbb S^3$, and for the standard compact picture of an induced representation on the boundary.

The same symmetric space $X$ also admits an unbounded realization on either sheet of the unit timelike hyperboloid:
\begin{align}\label{eq:hyperbolic-sheets}
    \mathbb H^4_\epsilon
    =
    \{Y\in\mathbb R^{1,4} \;\colon\;\; Y\cdot Y=1,\ \epsilon Y^0>0\}
    \,, \qquad
    \epsilon=\pm1 \,.
\end{align}
The connected group $G$ preserves each sheet and acts transitively on it, with stabilizer isomorphic to $K$. For the upper sheet, the identification with the ball model is
\begin{align}\label{eq:ball-hyperboloid-map}
    Y^0=\frac{1+|z|^2}{1-|z|^2}
    \,, \qquad
    \boldsymbol Y=\frac{2z}{1-|z|^2}
    \,, \qquad
    z=\frac{\boldsymbol Y}{Y^0+1} \,.
\end{align}
This map is $G$-equivariant when the ball is equipped with the action induced from the hyperboloid. The antipodal map $Y\mapsto-Y$ gives an equivariant identification with the lower sheet. The sheet label $\epsilon$ is independent of the representation label $\pm$ in $\Pi^\pm_{p,q}$.

Each sheet carries the Riemannian hyperbolic metric obtained by reversing the sign of the restricted ambient metric. Thus, both describe $\mathbb H^4$, whereas the Lorentzian spacetime $\mathrm{dS}_{4}$ is defined by the distinct hyperboloid $x\cdot x=-R^2$. The ideal boundary of the upper sheet consists of future-directed null rays and is diffeomorphic to $\mathbb S^3$; the lower sheet has the corresponding past-directed boundary, identified with it by the antipodal map. Each model therefore has a single ideal boundary $\mathbb S^3$.

Another unbounded realization of $X$ is the upper half-space model,
\begin{align}
    \mathbb U^4
    =\{(\xi,t) \;\colon\;\; \xi\in\mathbb R^3,\ t>0\}
    \,, \qquad
    ds^2=\frac{d\xi^2+dt^2}{t^2} \,.
\end{align}
Its ideal boundary is the one-point compactification $\mathbb R^3\cup\{\infty\}\simeq\mathbb S^3$, with $\mathbb R^3$ providing a stereographic chart on the sphere. The ball, hyperboloid sheet, and upper half-space therefore describe the same hyperbolic geometry and its ideal boundary. Passing between them changes the geometric model of the realization; it does not introduce an additional bulk--boundary correspondence.

For a non-scalar discrete-series UIR $\Pi^\pm_{p,q}$, let $\tau^\pm_{p,q}$ denote its minimal $K$-type, acting on $V_{\tau^\pm_{p,q}}$. This determines the homogeneous vector bundle
\begin{align}\label{eq:compact-functions}
    E_{p,q}^{\pm}
    =G\times_K V_{\tau_{p,q}^{\pm}}
    \longrightarrow X \,.
\end{align}
The representation is realized on a distinguished subspace of square-integrable sections of this bundle, selected by a $G$-equivariant differential system. In the Schmid realization, this system is expressed through a first-order differential operator $D^\pm_{p,q}$ \cite{KnappWallach1976, KnappWallach1980, Seppanen2009}. We write
\begin{align}\label{eq:compact-Hilbert-space}
    \mathcal H_{p,q}^{\pm}(X)
    =
    \left\{
        f\in L^2(X,E_{p,q}^{\pm})
        \;\colon\;\; D_{p,q}^{\pm}f=0
    \right\},
\end{align}
where the equation is understood in the distributional sense, so that the kernel is a closed Hilbert subspace. The action of $G$ on sections restricts to this subspace and carries the UIR $\Pi^\pm_{p,q}$.

The defining differential constraints are essential: analyticity alone does not isolate the desired representation. Moreover, although the real four-ball can be identified with the unit ball in $\mathbb C^2$ as a real manifold, the symmetric space $G/K$ admits no $G$-invariant complex structure. The bounded realization should therefore be understood as a space of constrained bundle sections, rather than an unrestricted space of holomorphic functions on a complex two-ball.

The invariant Hermitian fibre metric $h_{p,q}^{\pm}$ defines the Hilbert inner product
\begin{align}\label{eq:compact-inner-product}
    \langle f_1,f_2\rangle_{p,q}
    =
    C_{p,q}\int_{\mathbb B}
    h_{p,q;z}^{\pm}\bigl(f_1(z),f_2(z)\bigr)\,d\nu(z)
    \,, \qquad
    d\nu(z)=\frac{d^4z}{(1-|z|^2)^4} \,,
\end{align}
where $C_{p,q}>0$ absorbs the constant factor in the hyperbolic volume element and fixes the overall normalization. The measure $d\nu$ is $G$-invariant and independent of the representation labels. The dependence on $(p,q)$ enters through the bundle, its fibre metric, and the differential constraints defining $\mathcal H^\pm_{p,q}$. In a chosen trivialization, the fibre metric may introduce additional weights, whose transformation is compensated by the bundle cocycle. Such weights are not separately invariant measures on $G/K$, and no universal scalar weight is assumed for all $(p,q)$.

For the strictly massless family, the minimal $K$-types are chiral. Writing them as representations of the covering group $\mathrm{Spin}(4)\simeq\mathrm{SU}(2)\times\mathrm{SU}(2)$, we adopt the sign convention
\begin{align}\label{eq:compact-K-types}
    \tau_{p,p}^{-}=\mathcal D^p\otimes\mathbbm1
    \,, \qquad
    \tau_{p,p}^{+}=\mathbbm1\otimes\mathcal D^p \,,
\end{align}
where $\mathcal D^p$ is the spin-$p$ representation of $\mathrm{SU}(2)$ and $\mathbbm1$ denotes the trivial representation. For integer $p$, these representations descend to $K=\mathrm{SO}(4)$. They specify the bundle fibres, each of dimension $2p+1$, rather than the full $K$-type decomposition of the infinite-dimensional UIR. For the partially massless representations with $q<p$, the corresponding minimal $K$-types $\tau^\pm_{p,q}$ must be used; the same chiral fibre cannot be assumed throughout the series.

The identification in Eq. \eqref{eq:ball-hyperboloid-map} transports the bundle sections, fibre metric, and defining differential system from the ball to the upper hyperboloid sheet. It therefore induces a unitary equivalence
\begin{align}\label{eq:model-transport}
    T:\mathcal H_{p,q}^{\pm}(\mathbb B)
    \xrightarrow{\simeq}
    \mathcal H_{p,q}^{\pm}(\mathbb H^4_+)
    \,, \qquad
    \Pi^{\mathbb H}(g)
    =
    T\Pi^{\mathbb B}(g)T^{-1}
    \,, \quad g\in G \,.
\end{align}
In the trivialization transported by this identification, the map is simply
\begin{align}
    (Tf)(Y)=f\bigl(z(Y)\bigr) \,.
\end{align}
This equivalence expresses the same bundle realization in two geometric models of $X$. Relating it to a covariant field realization on $\mathrm{dS}_{4}$ requires a separate intertwining construction involving the physical gauge quotient.

%%%%%%%%%%%%%%%%%%%%%%%%%%%%%%%%%%%%%%%%%%%%%%%%%%%%%
%%%%%%%%%%%%%%%%%%%%%%%%%%%%%%%%%%%%%%%%%%%%%%%%%%%%%
\subsection{Coadjoint Orbits and the Meaning of Singular Sectors} \label{subsec:orbits}

The symmetric space $X=G/K$ serves as the base of the analytic bundle realization described above. This role must be distinguished from that of a classical phase space. A coadjoint orbit through $F\in\mathfrak g^\star$ is the homogeneous space
\begin{align}
    \mathcal O_F=G/G_F \,,
\end{align}
where $G_F$ is the stabilizer of $F$. It carries the Kirillov--Kostant--Souriau symplectic form
\begin{align}
    \omega_F\bigl(\operatorname{ad}^\star_A F \;,\;
                 \operatorname{ad}^\star_B F\bigr)
    =\langle F \;,\; [A,B]\rangle
    \,, \qquad A,B\in\mathfrak g \,.
\end{align}
Under the identification $\mathfrak g^\star\simeq\mathfrak g$ induced by the Killing form, $G_F$ becomes the centralizer of the corresponding Lie algebra element.

The space $G/K$ cannot itself be a coadjoint orbit with its natural $G$-action. Indeed, in the Cartan decomposition $\mathfrak g=\mathfrak k\oplus\mathfrak p$, the $K$-fixed subspace is
\begin{align}
    \mathfrak g^K
    =
    \mathfrak z(\mathfrak k)\oplus\mathfrak p^K
    =\{0\} \,.
\end{align}
Thus, the only coadjoint element fixed by all of $K$ is zero, whose orbit is a point. Equivalently, the isotropy representation of $K=\mathrm{SO}(4)$ on $T_oX\simeq\mathbb R^4$ admits no non-zero invariant alternating two-form, and hence no invariant symplectic form.

The relation between the symmetric space and coadjoint orbits can nevertheless be made explicit. Let $J_{12}$ and $J_{34}$ denote real Lie algebra generators of rotations in two orthogonal spatial planes, and use the Killing identification to write
\begin{align}
    F_{a,b}=aJ_{12}+bJ_{34} \,.
\end{align}
The corresponding centralizers give
\begin{align}\label{eq:elliptic-orbits}
    a>b>0\;\colon\;\; \quad&
    G_{F_{a,b}}\simeq\mathrm{SO}(2)\times\mathrm{SO}(2) \,,
    &\dim\mathcal O_{F_{a,b}}=&8 \,,
     \\[3pt]
    a=b>0\;\colon\;\; \quad&
    G_{F_{a,a}}\simeq\mathrm U(2) \,,
    &\dim\mathcal O_{F_{a,a}}=&6 \,.
\end{align}
The opposite-chirality choice $F_{a,-a}$ yields an analogous six-dimensional orbit, with the other embedding of $\mathrm U(2)$ in $K$.

Since these stabilizers lie in $K$, the orbits admit the fibrations
\begin{align}
    K/G_F \longrightarrow G/G_F \longrightarrow G/K \,.
\end{align}
The compact fibres have dimensions four and two, respectively; in the equal-rotation case, $K/G_F\simeq\mathbb S^2$. The ball therefore describes the base of the fibration, while the symplectic phase space is its total space. Since $G/K$ is contractible, each orbit has the homotopy type of its compact fibre. In particular, these orbits are not homeomorphic to the open ball.

In the classical particle-orbit parametrization of Ref. \cite{BasileJoungOh2024}, equal non-zero rotation parameters describe strictly massless spinning particles, while appropriate quantized unequal parameters describe partially massless particles. The latter remain regular eight-dimensional elliptic orbits in four dimensions: partial masslessness does not entail the stabilizer enhancement found in the strictly massless case. The six-dimensional equal-rotation orbit is called singular because its stabilizer is larger and its dimension smaller than those of a regular orbit. It remains a smooth symplectic manifold.

This classical notion of singularity must also be distinguished from singularity of a quantum infinitesimal character. The classical parameters $(a,b)$ cannot be identified directly with the Dixmier labels $(p,q)$ without specifying the quantization prescription and its shifts. In orthonormal $B_2$ coordinates, a representative of the infinitesimal character of $\Pi^\pm_{p,q}$ is
\begin{align}\label{eq:HC-shift}
    \lambda_{p,q}
    =
    \left(p+\frac12,q-\frac12\right)
    \,, \qquad
    \langle Q^{(1)}\rangle
    =
    \frac52-|\lambda_{p,q}|^2 \,.
\end{align}
The product over the positive roots is proportional to
\begin{align}
    \lambda_1\lambda_2
    (\lambda_1-\lambda_2)(\lambda_1+\lambda_2) \,.
\end{align}
It is non-zero for $p\geqslant q\geqslant1$, including $p=q$. Thus, the strictly massless genuine discrete series has a regular infinitesimal character. The reduction in dimension of the classical massless particle orbit and the regularity of the quantum infinitesimal character concern different objects.

The exceptional scalar modules $\Pi_{p,0}$ require a separate analysis. Their finite-dimensional gauge sectors and physical quotients do not by themselves determine a coadjoint-orbit realization. Nor does $q=0$ imply a singular infinitesimal character: the shifted coordinates $\lambda_{p,0}=(p+\tfrac12,-\tfrac12)$ remain regular for $p\geqslant1$. In fact, $\lambda_{p,0}$ and $\lambda_{p,1}$ are related by a Weyl reflection, so the corresponding modules have the same infinitesimal character and, in particular, the same two Casimir eigenvalues, despite being inequivalent representations. Any proposed orbit construction for the scalar family must therefore specify the orbit, polarization, and reduction explicitly; these data cannot be inferred from the Casimir eigenvalues alone.

%%%%%%%%%%%%%%%%%%%%%%%%%%%%%%%%%%%%%%%%%%%%%%%%%%%%%
%%%%%%%%%%%%%%%%%%%%%%%%%%%%%%%%%%%%%%%%%%%%%%%%%%%%%
\subsection{Covariant Spacetime Realization on $\mathrm{dS}_{4}$} \label{subsec:noncompact-realization}

We now describe the covariant spacetime realization, following Ref. \cite{Gazeau2024}. The spacetime carrier is the one-sheeted hyperboloid
\begin{align} \label{eq:dS-hyperboloid}
    \mathrm{dS}_{4}
    :=
    \left\{
        x=(x^a)\in\mathbb R^{1,4}
        \;\colon\;\;
        x\cdot x=\eta_{ab}x^ax^b=-R^2
    \right\} \,,
\end{align}
where $\eta_{ab}=\operatorname{diag}(1,-1,-1,-1,-1)$ and $R>0$ is the curvature radius. Its homogeneous-space description is $\mathrm{dS}_{4}\simeq G/\mathrm{SO}_0(1,3)$, and its topology is $\mathbb R\times\mathbb S^3$. Equipped with the induced Lorentzian metric, it provides the spacetime on which the covariant fields are defined. This carrier is distinct from both the Riemannian symmetric space $G/K$ and the coadjoint phase spaces discussed above.

We restrict attention to the integer-spin fields relevant to our program. A convenient ambient description uses symmetric rank-$r$ tensor fields $\psi^{(r)}_{a_1\cdots a_r}(x)$ satisfying tangentiality in every index:
\begin{align}
    x^{a_i}\psi^{(r)}_{a_1\cdots a_i\cdots a_r}(x)=0
    \,, \qquad i=1,\cdots,r \,.
\end{align}
This condition ensures that the ambient tensor defines an intrinsic tensor field on the hyperboloid. The infinitesimal group action is generated by
\begin{align} \label{eq:ambient-generators}
    L^{(r)}_{ab}=M^{}_{ab}+S^{(r)}_{ab} \,,
\end{align}
where the orbital part is
\begin{align}
    M_{ab}
    =
    -\mathrm{i}\left(x_a\partial_b-x_b\partial_a\right) \,,
\end{align}
and the spin part acts on the tensor indices as
\begin{align}
    \bigl(S^{(r)}_{ab}\psi^{(r)}\bigr)_{a_1\cdots a_r}
    =
    -\mathrm{i}\sum_{i=1}^{r}
    \left(
        \eta_{aa_i}
        \psi^{(r)}_{a_1\cdots a_{i-1}b a_{i+1}\cdots a_r}
        -
        \eta_{ba_i}
        \psi^{(r)}_{a_1\cdots a_{i-1}a a_{i+1}\cdots a_r}
    \right) \,.
\end{align}
Here, the derivatives act on an ambient extension of the field; the orbital combinations above are tangent to the hyperboloid. The total generators satisfy the commutation relations in Eq. \eqref{eq:so14-commutation}, preserve tangentiality, and commute with $x\cdot x$. Their action therefore restricts consistently to tensor fields on $\mathrm{dS}_{4}$.

The Casimir operators in Eq. \eqref{eq:discrete-casimirs-def} consequently take the rank-$r$ form
\begin{align}
    Q^{(1)}_r
    =&-\frac12 L^{(r)}_{ab}L^{(r)ab}
    \,, \qquad
    Q^{(2)}_r=-W^{(r)}_aW^{(r)a} \,,
     \\[3pt]
    W^{(r)}_a
    =&-\frac18\epsilon_{abcde}
    L^{(r)bc}L^{(r)de} \,.
\end{align}
Their joint eigenvalues constrain the representation content but do not uniquely identify the desired UIR. The quadratic Casimir provides a natural starting point for the field equations because it is a second-order differential operator. For the labels $(p,q)$, its eigenvalue equation is
\begin{align} \label{eq:spacetime-Casimir-equation}
    \left(
        Q^{(1)}_r+p(p+1)+(q+1)(q-2)
    \right)\psi^{(r)}=0 \,.
\end{align}
This equation must be supplemented by the conditions appropriate to the chosen tensor realization. In gauge-field descriptions, it may arise after imposing subsidiary conditions on a larger gauge-invariant system. The full covariant solution space can therefore contain additional invariant sectors, and identifying the physical UIR generally requires a quotient by gauge solutions rather than merely selecting an invariant subspace.

The three examples considered below are the strictly massless vector associated with $\Pi^\pm_{1,1}$, the graviton associated with $\Pi^\pm_{2,2}$, and the partially massless spin-$2$ field associated with $\Pi^\pm_{2,1}$. Their covariant gauge-field realizations carry indecomposable representations, with the physical UIRs recovered through the corresponding Gupta--Bleuler reductions. This distinguishes them from the bounded realization on $\mathcal H^\pm_{p,q}(\mathbb B)$, where the UIR is carried directly by a positive-definite Hilbert space. The examples below make the relevant invariant subspaces and physical quotients explicit.

For the spacetime analysis, we use the global conformal coordinates
\begin{align}\label{coor}
    x=x(\rho,u)
    =
    \left(R\tan\rho \,,\, R\sec\rho\,u\right)
    \,, \qquad
    u\in\mathbb S^3
    \,, \qquad
    -\frac{\pi}{2}<\rho<\frac{\pi}{2} \,.
\end{align}
The induced metric is
\begin{align}
    ds^2
    =
    R^2\sec^2\rho
    \left(d\rho^2-d\Omega_3^2\right) \,,
\end{align}
where $d\Omega_3^2$ denotes the round metric on the unit three-sphere, parametrized by $u\in\mathbb S^3$, and $\mathrm{d}u$ is its volume element. Each constant-$\rho$ slice is a global Cauchy surface; the slice $\rho=0$ is particularly convenient for defining the invariant field inner products. The limits $\rho\to\pm\pi/2$ correspond to the future and past conformal boundaries $\mathscr I^\pm\simeq\mathbb S^3$, respectively.

%%%%%%%%%%%%%%%%%%%%%%%%%%%%%%%%%%%%%%%%%%%%%%%%%%%%%
%%%%%%%%%%%%%%%%%%%%%%%%%%%%%%%%%%%%%%%%%%%%%%%%%%%%%
\subsubsection{Spacetime Realization: the Spin-$1$ Discrete Series} \label{sec:vector-spacetime}

We consider the strictly massless spin-$1$ representations $\Pi_{1,1}^{\pm}$, following Ref. \cite{takVec}. The construction proceeds from the Casimir equation to the covariant Maxwell equation, whose physical solution space is obtained by Gupta--Bleuler reduction.

Let $K_a(x):=\psi^{(1)}_a(x)$ be a vector field on the $\mathrm{dS}_{4}$ hyperboloid, represented by a homogeneous ambient extension and satisfying
\begin{align}
    x\cdot K=0 \,.
\end{align}
This transversality condition ensures that $K$ is tangent to spacetime. With
\begin{align}
    \partial^{\top}_a
    :=\theta_{ab}\partial^b
    =\partial_a+R^{-2}x_a x\cdot\partial \,,
    \qquad
    \theta_{ab}:=\eta_{ab}+R^{-2}x_a x_b \,,
\end{align}
the quadratic Casimir acts as
\begin{align}\label{Q1}
    \bigl(Q^{(1)}_1K\bigr)_a
    =&(Q^{(1)}_0-2)K_a
      +2x_a\partial^{\top}\cdot K \,,
    \qquad
    Q^{(1)}_0=-R^2\square_{\mathrm{dS}_{4}}
             =-R^2\, \big(\partial^{\top}\big)^2 \,.
\end{align}
Here, $Q^{(1)}_0$ acts on each ambient component of $K$.

For $p=q=1$, the Casimir eigenvalue is zero. Together with the divergencelessness condition, this gives 
\begin{align}\label{eq:vector-constrained-system}
    Q^{(1)}_1K=0 \,,
    \qquad
    \partial^{\top}\cdot K=0 \,.
\end{align}
The first condition fixes the quadratic Casimir, while the second removes the divergence degree of freedom. Neither condition, however, removes the pure-gauge solutions. Indeed, the tangential gradient
\begin{align}
    D_1:=R^2\partial^{\top}
\end{align}
satisfies
\begin{align}\label{eq:vector-gauge-identities}
    Q^{(1)}_1D_1\phi=D_1Q^{(1)}_0\phi \,,
    \qquad
    \partial^{\top}\cdot D_1\phi= - Q^{(1)}_0\phi \,.
\end{align}
Hence, every field $K_{\rm g}=D_1\phi_{\rm g}$ with $Q^{(1)}_0\phi_{\rm g}=0$ solves Eq. \eqref{eq:vector-constrained-system}. The constrained system therefore retains the freedom
\begin{align}\label{eq:vector-residual-gauge}
    K\longmapsto K+D_1\phi_{\rm g} \,,
    \qquad
    Q^{(1)}_0\phi_{\rm g}=0 \,.
\end{align}
Its physical states must be defined modulo these pure-gauge solutions.

To formulate the dynamics before imposing divergencelessness, consider
\begin{align}
    \mathcal M K
    :=Q^{(1)}_1K+D_1\partial^{\top}\cdot K \,.
\end{align}
Equation \eqref{eq:vector-gauge-identities} gives the operator identity
\begin{align}
    \mathcal M D_1\phi
    =D_1Q^{(1)}_0\phi-D_1Q^{(1)}_0\phi
    =0 \,.
\end{align}
Thus, the equation
\begin{align}\label{eq:vector-gauge-invariant}
    Q^{(1)}_1K+D_1\partial^{\top}\cdot K=0
\end{align}
is invariant under $K\mapsto K+D_1\phi$ for an arbitrary smooth scalar parameter $\phi$. It is the source-free Maxwell equation in ambient notation. Imposing $\partial^{\top}\cdot K=0$ recovers Eq. \eqref{eq:vector-constrained-system}, and the transformations preserving this condition are precisely those in Eq. \eqref{eq:vector-residual-gauge}. The divergenceless Casimir system is therefore the Lorenz-gauge form of the Maxwell equation.

For the covariant Gupta--Bleuler construction, one uses the gauge-fixed equation
\begin{align}\label{eq:vector-gauge-fixed}
    Q^{(1)}_1K+cD_1\partial^{\top}\cdot K=0 \,,
    \qquad c\in\mathbb R \,.
\end{align}
The value $c=1$ gives the gauge-invariant equation. For $c\neq1$, the identity $\partial^{\top}\cdot Q^{(1)}_1K =Q^{(1)}_0(\partial^{\top}\cdot K)$ implies
\begin{align}
    (1-c)Q^{(1)}_0(\partial^{\top}\cdot K)=0 \,.
\end{align}
Consequently, the divergence data and the gauge parameters of the divergenceless pure-gauge solutions obey
\begin{align}\label{eq:vector-scalar-sectors}
    Q^{(1)}_0(\partial^{\top}\cdot K)=0 \,,
    \qquad
    Q^{(1)}_0\phi_{\rm g}=0 \,.
\end{align}
Both are governed by the minimally coupled scalar equation. Their relation to the exceptional scalar representation $\Pi_{1,0}$ will enter the description of the non-physical sectors.

The invariant sesquilinear form on the covariant mode space is \cite{takVec}
\begin{align}\label{eq:vector-inner-product}
    \langle K_1,K_2\rangle_c
    ={}&\mathrm{i}R^2
    \int_{\mathbb S^3,\;\rho=0}
    \Big[
        {K_1^\ast}\cdot\partial_\rho K^{}_2
        -K^{}_2\cdot\partial_\rho{K_1^\ast}
        \nonumber \\[3pt]
        &\qquad
        -c\bigl((\partial_\rho x)\cdot{K_1^\ast}\bigr)
          \bigl(\partial^{\top}\cdot K^{}_2\bigr)
        +c\bigl((\partial_\rho x)\cdot K^{}_2\bigr)
          \bigl(\partial^{\top}\cdot{K_1^\ast}\bigr)
    \Big]\,\mathrm{d}u \,.
\end{align}
Its restriction to divergenceless fields is independent of $c$:
\begin{align}
    \langle K_1,K_2\rangle_{\rm KG}
    =\mathrm{i}R^2
    \int_{\mathbb S^3,\;\rho=0}
    \left[
        {K_1^\ast}\cdot\partial_\rho K^{}_2
        -K^{}_2\cdot\partial_\rho{K_1^\ast}
    \right]\,\mathrm{d}u \,.
\end{align}

Let $V_{\rm c}^{\rm vec}$ denote the covariant one-particle mode space of Eq. \eqref{eq:vector-gauge-fixed}, with $c\neq1$, as constructed in Ref. \cite{takVec}. Its invariant subspaces are
\begin{align}
    V^{\rm vec}
    :=&
    \left\{
        K\in V_{\rm c}^{\rm vec}
        \;\colon\;\;
        \partial^{\top}\cdot K=0
    \right\} \,,
     \\[3pt]
    V_{\rm g}^{\rm vec}
    :=&
    \left\{
        K=D_1\phi_{\rm g}\in V^{\rm vec}
        \;\colon\;\;
        Q^{(1)}_0\phi_{\rm g}=0
    \right\} \,.
\end{align}
They form the Gupta--Bleuler triplet
\begin{align}
    V_{\rm g}^{\rm vec}
    \subset V^{\rm vec}
    \subset V_{\rm c}^{\rm vec} \,.
\end{align}
The form \eqref{eq:vector-inner-product} is indefinite on $V_{\rm c}^{\rm vec}$. On $V^{\rm vec}$ it is positive semidefinite, and its radical is precisely $V_{\rm g}^{\rm vec}$:
\begin{align}
    \langle K_{\rm g},K\rangle_c=0 \,,
    \qquad
    \text{for all}\quad
    K_{\rm g}\in V_{\rm g}^{\rm vec} \,,
    \quad K\in V^{\rm vec} \,.
\end{align}
The physical inner product is therefore defined on gauge-equivalence classes by
\begin{align}
    \langle[K_1],[K_2]\rangle_{\rm phys}
    :=\langle K_1,K_2\rangle_c \,.
\end{align}
It is independent of the representatives and positive definite. The resulting physical Hilbert space, with completion understood, is
\begin{align}
    {\cal H}_{\rm phys}^{\rm vec}
    =V^{\rm vec}/V_{\rm g}^{\rm vec} \,,
    \qquad
    \Pi_{\rm phys}^{\rm vec}
    \simeq\Pi_{1,1}^{+}\oplus\Pi_{1,1}^{-} \,.
\end{align}
Thus, the two-photon helicity UIRs occur on the quotient, rather than on the full divergenceless space.

The covariant representation is indecomposable: $V^{\rm vec}$ has no invariant complement in $V_{\rm c}^{\rm vec}$, and $V_{\rm g}^{\rm vec}$ has no invariant complement in $V^{\rm vec}$. Its successive sectors are summarized by
\begin{align}
    \underbrace{
        V_{\rm c}^{\rm vec}/V^{\rm vec}
    }_{\text{divergence data}}
    \;\longrightarrow\;
    \underbrace{
        V^{\rm vec}/V_{\rm g}^{\rm vec}
    }_{\Pi_{1,1}^{+}\oplus\Pi_{1,1}^{-}}
    \;\longrightarrow\;
    \underbrace{
        V_{\rm g}^{\rm vec}
    }_{\text{pure-gauge fields}} \,.
\end{align}
The arrows indicate the non-trivial extensions under the $\mathrm{SO}_0(1,4)$ action, not maps between these subquotients. Equation \eqref{eq:vector-scalar-sectors} explains the scalar character of the outer sectors. Their customary association with $\Pi_{1,0}$ suppresses the finer minimally coupled scalar structure, including its constant mode and Gupta--Bleuler reduction \cite{takVec}.

For the holographic program, the relevant spacetime carrier is ${\cal H}_{\rm phys}^{\rm vec}$. A boundary construction must therefore map pure-gauge fields to physically trivial boundary data and descend to an intertwiner on this quotient, preserving its positive-definite inner product.

%%%%%%%%%%%%%%%%%%%%%%%%%%%%%%%%%%%%%%%%%%%%%%%%%%%%%
%%%%%%%%%%%%%%%%%%%%%%%%%%%%%%%%%%%%%%%%%%%%%%%%%%%%%
\subsubsection{Spacetime Realization: the Graviton Discrete Series} \label{sec:graviton-spacetime}

We now consider the strictly massless spin-$2$ representations $\Pi_{2,2}^{\pm}$, following Ref. \cite{GravGB}. As in the vector case, their covariant spacetime realization contains ordinary pure-gauge fields that must be removed to obtain the physical carrier. For the graviton, however, this reduction is intertwined with a second structure: the minimally coupled scalar function underlying the tensor modes introduces a constant zero mode and an associated gauge-like ambiguity. The physical Hilbert space is therefore obtained through two successive reductions within the same covariant field space.

Let ${\cal K}_{ab}(x):=\psi^{(2)}_{ab}(x)$ be a symmetric rank-$2$ tensor field on $\mathrm{dS}_{4}$, represented by a homogeneous ambient extension and satisfying
\begin{align}
    x^a{\cal K}_{ab}=0 \,.
\end{align}
On such fields, the quadratic Casimir acts as
\begin{align}
    \bigl(Q^{(1)}_2{\cal K}\bigr)_{ab}
    =&
    (Q^{(1)}_0-6){\cal K}_{ab}
    +2{\cal S}\bigl(x_a\partial^{\top}\cdot{\cal K}_{\cdot b}\bigr)
    +2\eta_{ab}{\cal K}^{\prime} \,,
\end{align}
where ${\cal K}^{\prime}:=\eta^{ab}{\cal K}_{ab}$ and ${\cal S}T_{ab}:=T_{ab}+T_{ba}$ denotes symmetrization without a factor of $1/2$. For $p=q=2$, the Casimir eigenvalue is $-6$, giving the
constrained system
\begin{align}\label{eq:graviton-constrained-system}
    (Q^{(1)}_2+6){\cal K}=0 \,,
    \qquad
    \partial^{\top}\cdot{\cal K}=0 \,,
    \qquad
    {\cal K}^{\prime}=0 \,.
\end{align}
Together with ambient transversality, these conditions select the divergenceless and traceless tensor sector. They do not, however, eliminate its pure-gauge solutions.

To describe this freedom, introduce the symmetrized tangential gradient
\begin{align}
    (D_2\Lambda)_{ab}
    :=&
    R^2\bigl(
        \partial^{\top}_a\Lambda_b
        +\partial^{\top}_b\Lambda_a
    \bigr)
    -x_a\Lambda_b-x_b\Lambda_a
\end{align}
and the generalized divergence
\begin{align}
    (\partial_2\cdot{\cal K})_b
    :=&
    \partial\cdot{\cal K}_{\cdot b}
    -R^{-2}x_b{\cal K}^{\prime}
    -\frac12\partial^{\top}_b{\cal K}^{\prime} \,.
\end{align}
For an ambient-transverse traceless tensor, $\partial_2\cdot{\cal K}=\partial^{\top}\cdot{\cal K}$. For a transverse vector $\Lambda$, these operators satisfy 
\begin{align}\label{eq:graviton-gauge-identities}
    (Q^{(1)}_2+6)D_2\Lambda
    =&D_2(Q^{(1)}_1+6)\Lambda \,,
     \\[3pt]
    \partial_2\cdot D_2\Lambda
    =&-(Q^{(1)}_1+6)\Lambda \,,
     \\[3pt]
    (D_2\Lambda)^{\prime}
    =&2R^2\partial^{\top}\cdot\Lambda \,.
\end{align}
Consequently, ${\cal K}_{\rm g}=D_2\Lambda_{\rm g}$ solves Eq. \eqref{eq:graviton-constrained-system} whenever
\begin{align}
    x\cdot\Lambda_{\rm g}=0 \,,
    \qquad
    \partial^{\top}\cdot\Lambda_{\rm g}=0 \,,
    \qquad
    (Q^{(1)}_1+6)\Lambda_{\rm g}=0 \,.
\end{align}
Such fields represent the ordinary spin-$2$ gauge ambiguity.

The identities in Eq. \eqref{eq:graviton-gauge-identities} also give
\begin{align}
    \Bigl[
        (Q^{(1)}_2+6)+D_2\partial_2\cdot
    \Bigr]D_2\Lambda=0 \,.
\end{align}
Thus, the constrained system admits the gauge-invariant extension
\begin{align}\label{eq:graviton-field-equation}
    (Q^{(1)}_2+6){\cal K}
    +D_2\partial_2\cdot{\cal K}=0 \,,
\end{align}
which is the ambient form of the linearized Einstein equation and is invariant under
\begin{align}\label{eq:graviton-ordinary-gauge}
    {\cal K}\longmapsto{\cal K}+D_2\Lambda \,.
\end{align}
Here, $\Lambda$ is an arbitrary transverse vector field; preserving tracelessness additionally requires $\partial^{\top}\cdot\Lambda=0$.

Introducing a real gauge-fixing parameter $c$, we work with
\begin{align}\label{eq:graviton-c-equation}
    (Q^{(1)}_2+6){\cal K}
    +cD_2\partial_2\cdot{\cal K}=0 \,,
    \qquad
    {\cal K}^{\prime}=0 \,.
\end{align}
The value $c=1$ recovers the gauge-invariant equation, with the preceding restriction on gauge transformations when tracelessness is retained. For $c\neq1$, taking the generalized divergence gives
\begin{align}
    (1-c)(Q^{(1)}_1+6)\partial_2\cdot{\cal K}=0 \,.
\end{align}
Hence, the divergence data and the parameters generating divergenceless pure-gauge fields obey the same auxiliary vector equation:
\begin{align}\label{eq:graviton-auxiliary-sectors}
    (Q^{(1)}_1+6)\partial_2\cdot{\cal K}=0 \,,
    \qquad
    (Q^{(1)}_1+6)\Lambda_{\rm g}=0 \,.
\end{align}

Let $V_{\rm c}^{\rm grav}$ denote the complete covariant mode space used in Ref. \cite{GravGB}. It carries the invariant sesquilinear form
\begin{align}
    \langle{\cal K}_1,{\cal K}_2\rangle_c
    ={}&\mathrm{i}R^2
    \int_{\mathbb S^3,\;\rho=0}
    \Big[
        {{\cal K}^\ast_1}\mathbin{\cdot\cdot}
        \partial_\rho{\cal K}^{}_2
        -{\cal K}^{}_2\mathbin{\cdot\cdot}
        \partial_\rho{{\cal K}^\ast_1}
        \nonumber \\[3pt]
        &\qquad
        -2c\bigl((\partial_\rho x)\cdot
        {{\cal K}^\ast_1}\bigr)\cdot
        \bigl(\partial_2\cdot{\cal K}^{}_2\bigr)
        \nonumber \\[3pt]
        &\qquad
        +2c\bigl((\partial_\rho x)\cdot
        {\cal K}^{}_2\bigr)\cdot
        \bigl(\partial_2\cdot{{\cal K}^\ast_1}\bigr)
    \Big]\,\mathrm{d}u \,,
\end{align}
where $\mathbin{\cdot\cdot}$ denotes contraction of both tensor indices. On divergenceless fields, the gauge-dependent terms vanish, leaving
\begin{align}
    \langle{\cal K}_1,{\cal K}_2\rangle_{\rm KG}
    =
    \mathrm{i}R^2
    \int_{\mathbb S^3,\;\rho=0}
    \left[
        {{\cal K}^\ast_1}\mathbin{\cdot\cdot}
        \partial_\rho{\cal K}^{}_2
        -{\cal K}^{}_2\mathbin{\cdot\cdot}
        \partial_\rho{{\cal K}^\ast_1}
    \right]\,\mathrm{d}u \,.
\end{align}

The ordinary gauge structure is expressed by the invariant filtration
\begin{align}\label{eq:graviton-GB1}
    V_{{\rm g}_1}^{\rm grav}
    \subset V_1^{\rm grav}
    \subset V_{\rm c}^{\rm grav} \,,
\end{align}
where
\begin{align}
    V_1^{\rm grav}
    :=&
    \left\{
        {\cal K}\in V_{\rm c}^{\rm grav}
        \;\colon\;
        \partial_2\cdot{\cal K}=0
    \right\} \,,
     \\[3pt]
    V_{{\rm g}_1}^{\rm grav}
    :=&
    \left\{
        D_2\Lambda_{\rm g}\in V_1^{\rm grav}
    \right\} \,.
\end{align}
Its successive sectors describe divergence data, divergenceless fields modulo ordinary gauge transformations, and ordinary pure-gauge fields. At this stage, however, the middle quotient cannot yet be identified with the physical Hilbert space: the scalar zero-mode structure must also be taken into account.

In the mode construction of Ref. \cite{GravGB}, the tensor field is obtained by applying a spin-$2$ polarization operator to a minimally coupled scalar structure function:
\begin{align}\label{eq:graviton-scalar-structure}
    {\cal K}_{ab}={\cal E}_{ab}\phi \,,
    \qquad
    Q^{(1)}_0\phi=0 \,.
\end{align}
Polarization labels are suppressed. The scalar equation admits the globally constant solution $\phi(x)=\chi$, which belongs to the $L=0$ sector of the spherical-harmonic expansion on $\mathbb S^3$. The shift
\begin{align}
    \phi\longmapsto\phi+\chi \,,
    \qquad
    \chi=\mathrm{const.} \,,
\end{align}
therefore preserves the scalar equation and induces
\begin{align}\label{eq:graviton-gauge-like}
    {\cal K}\longmapsto{\cal K}+{\cal E}\chi \,.
\end{align}
The resulting gauge-like modes generate an invariant null sector $V_{{\rm g}_2}^{\rm grav}$. This sector is distinct from the ordinary pure-gauge sector $V_{{\rm g}_1}^{\rm grav}$ and introduces an additional ambiguity in the tensor realization.

To describe its role, let ${\cal V}_2^{\rm grav;+}$ denote the graviton mode space constructed from the $L\geqslant1$ scalar modes in the positive-frequency family used in Ref. \cite{GravGB}, where positive frequency is defined with respect to the global conformal time $\rho$. This space is not invariant on its own: the $\mathrm{dS}_{4}$ action can generate contributions from the constant zero mode. Adjoining the corresponding gauge-like sector gives the invariant space
\begin{align}
    V_2^{\rm grav}
    =
    {\cal V}_2^{\rm grav;+}
    \oplus V_{{\rm g}_2}^{\rm grav} \,.
\end{align}
The direct sum here is a vector-space decomposition; the positive-frequency summand is not separately invariant. Since $V_{{\rm g}_2}^{\rm grav}$ is invariant, the group action nevertheless descends to $V_2^{\rm grav}/V_{{\rm g}_2}^{\rm grav}$, whose classes can be represented by modes in ${\cal V}_2^{\rm grav;+}$.

The form restricted to $V_2^{\rm grav}$ is degenerate, because its gauge-like modes are orthogonal to every element of this space. A non-degenerate covariant extension must therefore contain modes that pair non-trivially with them. In the scalar construction, these arise from the second, time-dependent solution in the $L=0$ sector. Writing ${\cal V}_{\rm s}^{\rm grav}$ for the corresponding tensor sector, one has
\begin{align}
    {\cal V}_{\rm s}^{\rm grav}
    \not\perp_{\rm KG}
    V_{{\rm g}_2}^{\rm grav} \,.
\end{align}
Adjoining this sector restores the missing pairing, but does not by itself give an invariant space: its $\mathrm{dS}_{4}$ orbit also generates negative-frequency modes. Including the resulting sector ${\cal V}_2^{\rm grav;-}$ leads to
\begin{align}
    V_{\rm c}^{\rm grav}
    =
    {\cal V}_2^{\rm grav;+}
    \oplus V_{{\rm g}_2}^{\rm grav}
    \oplus{\cal V}_2^{\rm grav;-}
    \oplus{\cal V}_{\rm s}^{\rm grav} \,.
\end{align}
Again, these summands are not separately invariant. The invariant structure is instead encoded by the second Gupta--Bleuler filtration,
\begin{align}\label{eq:graviton-GB2}
    V_{{\rm g}_2}^{\rm grav}
    \subset V_2^{\rm grav}
    \subset V_{\rm c}^{\rm grav} \,.
\end{align}
The time-dependent zero-mode partner and the negative-frequency sector thus enter for complementary reasons: the former supplies the missing pairing, while the latter is required for closure under the group action.

The physical reduction can now be stated precisely. First, remove the gauge-like null sector by setting
\begin{align}
    {\cal W}^{\rm grav}
    :=
    V_2^{\rm grav}/V_{{\rm g}_2}^{\rm grav} \,,
    \qquad
    q_2 \colon\,
    V_2^{\rm grav}\longrightarrow{\cal W}^{\rm grav} \,.
\end{align}
Because $V_{{\rm g}_2}^{\rm grav}$ is orthogonal to $V_2^{\rm grav}$, the invariant form descends to this quotient. The ordinary gauge structure survives in the invariant subspaces
\begin{align}
    {\cal W}_1^{\rm grav}
    :=&
    q_2\bigl(V_1^{\rm grav}\cap V_2^{\rm grav}\bigr) \,,
     \\[3pt]
    {\cal W}_{{\rm g}_1}^{\rm grav}
    :=&
    q_2\bigl(
        V_{{\rm g}_1}^{\rm grav}\cap V_2^{\rm grav}
    \bigr) \,.
\end{align}
Thus, ${\cal W}_1^{\rm grav}$ consists of classes admitting divergenceless representatives, whereas ${\cal W}_{{\rm g}_1}^{\rm grav}$ consists of classes represented by ordinary pure-gauge fields. They form the triplet
\begin{align}
    {\cal W}_{{\rm g}_1}^{\rm grav}
    \subset{\cal W}_1^{\rm grav}
    \subset{\cal W}^{\rm grav} \,.
\end{align}
The quotient ${\cal W}^{\rm grav}/{\cal W}_1^{\rm grav}$ describes the remaining divergence sector.

On ${\cal W}_1^{\rm grav}$, the induced invariant form is positive semidefinite, with radical ${\cal W}_{{\rm g}_1}^{\rm grav}$. Removing this remaining null sector and completing in the induced positive-definite norm gives
\begin{align}\label{eq:graviton-physical-space}
    {\cal H}_{\rm phys}^{\rm grav}
    :=&
    \overline{
        {\cal W}_1^{\rm grav}/
        {\cal W}_{{\rm g}_1}^{\rm grav}
    } \,,
     \\[3pt]
    \Pi_{\rm phys}^{\rm grav}
    \simeq&
    \Pi_{2,2}^{+}\oplus\Pi_{2,2}^{-} \,.
\end{align}
A physical graviton state is therefore represented by a divergenceless tensor mode, modulo both the scalar zero-mode ambiguity and the ordinary spin-$2$ gauge ambiguity.

Following Ref. \cite{GravGB}, denote by $\pi_{2,2}$ the auxiliary module labelled there by $(2,-1)$, which is Weyl related to $\Pi_{2,2}^{\pm}$. The ordinary gauge structure after the first reduction takes the form
\begin{align}\label{eq:graviton-basic-indec}
    \underbrace{\pi_{2,2}}_{
        {\cal W}^{\rm grav}/{\cal W}_1^{\rm grav}}
    \;\longrightarrow\;
    \underbrace{\Pi_{2,2}^{+}\oplus\Pi_{2,2}^{-}}_{
        {\cal W}_1^{\rm grav}/{\cal W}_{{\rm g}_1}^{\rm grav}}
    \;\longrightarrow\;
    \underbrace{\pi_{2,2}}_{
        {\cal W}_{{\rm g}_1}^{\rm grav}} \,.
\end{align}
The entries describe the divergence, physical, and ordinary pure-gauge sectors, respectively, with Hilbert completion understood for the physical quotient. The arrows encode non-trivial extensions under the group action; they do not denote linear maps between the displayed quotient spaces.

The corresponding extensions in the other layers of the zero-mode filtration combine with Eq. \eqref{eq:graviton-basic-indec} into the schematic diagram
\begin{align}\label{eq:graviton-complete}
\begin{array}{c@{\quad}c@{\quad}c@{\quad}c@{\quad}c@{\quad}c}
    V_{{\rm g}_2}^{\rm grav}\colon
    &\widetilde{\Upsilon}_0
    &\longrightarrow
    &\Upsilon_0
    &\longrightarrow
    &\widetilde{\Upsilon}_0
     \\[3pt]
    &\uparrow&&\uparrow&&\uparrow
     \\[3pt]
    {\cal W}^{\rm grav}\colon
    &\pi_{2,2}
    &\longrightarrow
    &\boxed{\Pi_{2,2}^{+}\oplus\Pi_{2,2}^{-}}
    &\longrightarrow
    &\pi_{2,2}
     \\[3pt]
    &\uparrow&&\uparrow&&\uparrow
     \\[3pt]
    V_{\rm c}^{\rm grav}/V_2^{\rm grav}\colon
    &\pi_{2,2}
    &\longrightarrow
    &\Pi_{2,2}^{+}\oplus\Pi_{2,2}^{-}
    &\longrightarrow
    &\pi_{2,2}
\end{array} \,.
\end{align}
Here, $\Upsilon_0$ denotes the trivial scalar representation, while $\widetilde{\Upsilon}_0$ labels the accompanying auxiliary zero-mode sectors in the notation of Ref. \cite{GravGB}. The rows correspond to the three layers of Eq. \eqref{eq:graviton-GB2}. Within each row, the horizontal arrows encode the ordinary spin-$2$ gauge extensions; the vertical arrows encode the extensions between the zero-mode layers. The displayed entries are therefore subquotients of the covariant representation, not invariant direct summands. 

The boxed entry locates the physical graviton carrier at the intersection of the two reductions: it is the middle quotient of the ordinary gauge filtration within the middle layer of the zero-mode filtration. A boundary construction must respect both structures so that it descends to ${\cal H}_{\rm phys}^{\rm grav}$ and intertwines its unitary $\mathrm{dS}_{4}$ action with that of the bounded realization, while preserving the corresponding Hilbert norms.

%%%%%%%%%%%%%%%%%%%%%%%%%%%%%%%%%%%%%%%%%%%%%%%%%%%%%
%%%%%%%%%%%%%%%%%%%%%%%%%%%%%%%%%%%%%%%%%%%%%%%%%%%%%
\subsubsection{Spacetime Realization: the Partially Massless Graviton Discrete Series} \label{sec:PM-spacetime}

We finally consider the partially massless spin-$2$ representations $\Pi^\pm_{2,1}$, following Ref. \cite{GravPartially}. Their covariant spacetime realization again involves a symmetric tensor field whose physical carrier is obtained by gauge reduction. In this case, however, the gauge freedom is generated by a scalar parameter, and the divergence and trace of the field are linked by a covariant constraint.

Let ${\cal K}_{ab}(x):=\psi^{(2)}_{ab}(x)$ be a symmetric rank-$2$ tensor field on $\mathrm{dS}_4$, represented by a homogeneous ambient extension and satisfying $x^a{\cal K}_{ab}=0$. For $p=2$ and $q=1$, the quadratic Casimir eigenvalue is $-4$, giving the constrained system
\begin{align}\label{eq:PM-constrained-system}
    (Q^{(1)}_2+4){\cal K}=0 \,,
    \qquad
    \partial^{\top}\cdot{\cal K}=0 \,,
    \qquad
    {\cal K}^{\prime}=0 \,.
\end{align}
As in the strictly massless spin-$2$ case, these conditions leave residual pure-gauge solutions. The physical representation is therefore carried by a quotient of this constrained sector.

To describe the larger covariant realization, one allows non-zero divergence and trace while retaining the relation
\begin{align}\label{eq:PM-constraint}
    \partial_2\cdot{\cal K}
    =
    \frac12\partial^{\top}{\cal K}^{\prime} \,,
\end{align}
where $\partial_2$ is the generalized divergence introduced in the graviton section. The gauge-invariant field equation takes the form
\begin{align}\label{eq:PM-field-equation}
    (Q^{(1)}_2+4){\cal K}
    +D_2\partial_2\cdot{\cal K}
    -\theta{\cal K}^{\prime}=0 \,,
\end{align}
with $D_1=R^2\partial^{\top}$ and $D_2={\cal S}(D_1-x)$, as above. Its gauge freedom is
\begin{align}\label{eq:PM-gauge-transformation}
    {\cal K}
    \longmapsto
    {\cal K}
    +D_2D_1\phi_{\rm g}
    -2R^2\theta\phi_{\rm g} \,.
\end{align}
Here, the scalar parameter $\phi_{\rm g}$ replaces the transverse vector parameter of the ordinary graviton gauge transformation.

Introducing a real gauge-fixing parameter $c$ gives
\begin{align}\label{eq:PM-gauge}
    (Q^{(1)}_2+4){\cal K}
    +cD_2\partial_2\cdot{\cal K}
    -c\theta{\cal K}^{\prime}=0 \,,
\end{align}
subject to Eq. \eqref{eq:PM-constraint}. The value $c=1$ recovers the gauge-invariant equation. On divergenceless and traceless fields, the $c$-dependent terms vanish, and one recovers Eq. \eqref{eq:PM-constrained-system}.

Let $V_{\rm c}^{\rm PMG}$ denote the covariant mode space of Eq. \eqref{eq:PM-gauge} for $c\neq1$, with Eq. \eqref{eq:PM-constraint} understood, as constructed in Ref. \cite{GravPartially}. It is equipped with the invariant sesquilinear form $\langle\cdot,\cdot\rangle_c$ introduced in the graviton section. Its relevant invariant subspaces are
\begin{align}
    V^{\rm PMG}
    :=&
    \left\{
        {\cal K}\in V_{\rm c}^{\rm PMG}
        \;\colon\;
        \partial_2\cdot{\cal K}=0,\quad
        {\cal K}^{\prime}=0
    \right\} \,,
    \\[3pt]
    V_{\rm g}^{\rm PMG}
    :=&
    \left\{
        D_2D_1\phi_{\rm g}
        -2R^2\theta\phi_{\rm g}
        \in V^{\rm PMG}
    \right\} \,.
\end{align}
The scalar parameters in the second line are restricted so that the resulting pure-gauge fields belong to the constrained solution space. These subspaces form the Gupta--Bleuler triplet
\begin{align}\label{eq:PM-GB}
    V_{\rm g}^{\rm PMG}
    \subset V^{\rm PMG}
    \subset V_{\rm c}^{\rm PMG} \,.
\end{align}
The outer quotient $V_{\rm c}^{\rm PMG}/V^{\rm PMG}$ describes the complementary scalar sector encoded by the trace and divergence, while $V^{\rm PMG}$ contains the physical modes together with their residual gauge ambiguity.

The invariant form is indefinite on $V_{\rm c}^{\rm PMG}$. On $V^{\rm PMG}$, it reduces to the $c$-independent form $\langle\cdot,\cdot\rangle_{\rm KG}$ and is positive semidefinite, with radical $V_{\rm g}^{\rm PMG}$. It therefore induces a positive-definite inner product on gauge-equivalence classes,
\begin{align}
    \bigl\langle[{\cal K}_1],[{\cal K}_2]\bigr\rangle_{\rm phys}
    :=
    \langle{\cal K}_1,{\cal K}_2\rangle_c \,.
\end{align}
Completing the quotient in the corresponding norm gives
\begin{align}\label{eq:PM-physical-space}
    {\cal H}_{\rm phys}^{\rm PMG}
    :=&
    \overline{
        V^{\rm PMG}/V_{\rm g}^{\rm PMG}
    } \,,
    \\[3pt]
    \Pi_{\rm phys}^{\rm PMG}
    \simeq&
    \Pi^+_{2,1}\oplus\Pi^-_{2,1} \,.
\end{align}
A physical partially massless graviton state is thus represented by a divergenceless and traceless tensor mode, modulo the scalar-generated pure-gauge fields.

For $c\neq1$, both the gauge sector and the complementary scalar sector are governed by the scalar discrete representation $\Pi_{2,0}$. Following Ref. \cite{GravPartially}, the corresponding indecomposable structure is summarized by
\begin{align}\label{eq:PM-indecomposable}
    \underbrace{\Pi_{2,0}}_{
        V_{\rm c}^{\rm PMG}/V^{\rm PMG}}
    \;\longrightarrow\;
    \underbrace{\Pi^+_{2,1}\oplus\Pi^-_{2,1}}_{
        V^{\rm PMG}/V_{\rm g}^{\rm PMG}}
    \;\longrightarrow\;
    \underbrace{\Pi_{2,0}}_{
        V_{\rm g}^{\rm PMG}} \,.
\end{align}
The entries describe the complementary scalar, physical, and pure-gauge sectors, respectively, with Hilbert completion understood for the physical quotient. The arrows encode non-trivial extensions under the $\mathrm{SO}_0(1,4)$ action; they do not denote linear maps between the displayed subquotients.

As in the vector case, the physical representation therefore occupies the central quotient of a larger indecomposable covariant representation. For the holographic construction, the boundary map must respect this gauge filtration: pure-gauge fields must give physically trivial boundary data, and the induced map on ${\cal H}_{\rm phys}^{\rm PMG}$ must intertwine the group actions and preserve the invariant Hilbert norm.

%%%%%%%%%%%%%%%%%%%%%%%%%%%%%%%%%%%%%%%%%%%%%%%%%%%%%
%%%%%%%%%%%%%%%%%%%%%%%%%%%%%%%%%%%%%%%%%%%%%%%%%%%%%
%%%%%%%%%%%%%%%%%%%%%%%%%%%%%%%%%%%%%%%%%%%%%%%%%%%%%
\section{Boundary Intertwiners: the Scalar Prototype and the Non-Scalar Bosonic Program} \label{subsec:compact-to-boundary}

The preceding spacetime constructions identify the physical carriers that the proposed boundary maps must recover. For the vector, graviton, and partially massless graviton, the physical representation emerges through the appropriate Gupta--Bleuler reduction of a larger covariant mode space. A spacetime boundary map must therefore respect gauge equivalence and descend to the physical quotient, while preserving its invariant Hilbert structure and intertwining the $\mathrm{SO}_0(1,4)$ actions.

The bounded and spacetime realizations lead to two distinct boundary problems. As described in Section \ref{subsec:compact-realization}, the symmetric space $X=G/K$ admits a bounded ball model $\mathbb B$ with ideal boundary $\partial\mathbb B\simeq\mathbb S^3$. The Lorentzian spacetime realization instead leads to asymptotic data on the conformal boundaries $\mathscr I^\pm\simeq\mathbb S^3$ of $\mathrm{dS}_4$. The common boundary geometry provides a setting in which the resulting representations can be compared, but does not by itself identify their carrier spaces, group actions, or invariant inner products. These structures must be constructed and related by explicit unitary intertwiners. 

On the symmetric-space side, the Knapp--Wallach Szeg\"o constructions provide an established starting point for relating discrete-series realizations to suitable boundary-induced representations \cite{KnappWallach1976, KnappWallach1980}. For the representations considered here, the task is to specify the relevant boundary data, constraints and quotients, determine the invariant Hilbert completion, and normalize the intertwining maps. The resulting boundary realization must then be matched to the one obtained from the physical spacetime quotient. A radial limit alone need not satisfy these requirements.

We begin with the exceptional scalar family $\Pi_{p,0}$, for which an explicit correspondence is already available \cite{Bajalan2026, bajalan2026araki}. Its physical spacetime realization admits an independent spherical realization directly on $\mathbb S^3$, and an explicit isometric intertwiner identifies their group actions and invariant Hilbert structures. The conformal-boundary analysis further relates this spherical carrier to asymptotic spacetime data. The scalar family thus provides a concrete prototype for combining boundary limits, gauge reduction, and unitary equivalence. We then formulate the additional boundary construction required for the non-scalar bosonic representations $\Pi^\pm_{p,q}$, whose bounded realizations are defined in the interior of $\mathbb B$.

%%%%%%%%%%%%%%%%%%%%%%%%%%%%%%%%%%%%%%%%%%%%%%%%%%%%%
%%%%%%%%%%%%%%%%%%%%%%%%%%%%%%%%%%%%%%%%%%%%%%%%%%%%%
\subsection{The Scalar Prototype: Spherical and Spacetime Realizations} \label{subsubsec:scalar-prototype}

Following Refs. \cite{Bajalan2026,bajalan2026araki}, we review the two realizations of $\Pi_{p,0}$, emphasizing the structures needed for their explicit identification. We first describe the covariant spacetime solution space and its Gupta--Bleuler quotient, then the spherical carrier and its invariant inner product. The conformal-boundary behavior of the spacetime modes connects these constructions, leading to the unitary intertwiner that establishes their equivalence.

%%%%%%%%%%%%%%%%%%%%%%%%%%%%%%%%%%%%%%%%%%%%%%%%%%%%%
%%%%%%%%%%%%%%%%%%%%%%%%%%%%%%%%%%%%%%%%%%%%%%%%%%%%%
\subsubsection{Spacetime Realization on $\mathrm{dS}_4$}

Let $\phi(x):=\psi^{(0)}(x)$ be a scalar field on $\mathrm{dS}_4$, represented by a homogeneous ambient extension. Since $L^{(0)}_{ab}=M_{ab}$, the quadratic Casimir is $Q^{(1)}_0=-R^2\square_{\rm dS}$. For $p\in\mathbb N$, the exceptional scalar representation $\Pi_{p,0}$ has eigenvalue
\begin{align}
    \langle Q^{(1)}_0\rangle=-(p-1)(p+2) \,,
\end{align}
and the corresponding field equation is
\begin{align}\label{eq:scalar-dS-wave}
    \left(
        R^2\square_{\rm dS}-(p-1)(p+2)
    \right)\phi=0 \,.
\end{align}
Its solutions carry the invariant Klein--Gordon sesquilinear form
\begin{align}
    \langle\phi_1,\phi_2\rangle_{\rm KG}
    =
    \mathrm{i}R^2
    \int_{\mathbb S^3,\;\rho=0}
    \left[
        \phi_1^\ast\,\partial_\rho\phi_2
        -\phi_2\,\partial_\rho\phi_1^\ast
    \right]\mathrm{d}u \,.
\end{align}

The covariant completion of the global mode space used in Refs. \cite{Bajalan2026,bajalan2026araki} is a non-degenerate Krein space,
\begin{align}
    V_{\rm tot}^{\rm sc}
    =
    {\cal H}_{+}^{\rm sc}\oplus{\cal H}_{-}^{\rm sc} \,,
    \qquad
    {\cal H}_{-}^{\rm sc}
    =
    \left\{
        \phi^\ast\;\colon\;\phi\in{\cal H}_{+}^{\rm sc}
    \right\} \,.
\end{align}
The Klein--Gordon form is positive definite on ${\cal H}_{+}^{\rm sc}$ and negative definite on ${\cal H}_{-}^{\rm sc}$; these summands are not separately invariant under the full $\mathrm{dS}_4$ action. The invariant structure relevant to the physical reduction is instead the Gupta--Bleuler triplet 
\begin{align}\label{GBScalar}
    V_{\rm g}^{\rm sc}
    \subset V^{\rm sc}
    \subset V_{\rm tot}^{\rm sc} \,.
\end{align}
Here, $V_{\rm g}^{\rm sc}$ is the finite-dimensional gauge sector, while the Klein--Gordon form restricted to $V^{\rm sc}$ is positive semidefinite, with radical $V_{\rm g}^{\rm sc}$.

The gauge sector has dimension
\begin{align}
    \dim V_{\rm g}^{\rm sc}
    =
    \frac{p(p+1)(2p+1)}{6} \,,
\end{align}
and carries the finite-dimensional irreducible representation $\pi_{p,0}$, labelled by $(0,p-1)$ and Weyl equivalent to $\Pi_{p,0}$. For $p>1$, this representation is non-unitary. For $p=1$, the gauge sector consists of constant solutions and carries the trivial representation.

Factoring out the radical makes the induced Klein--Gordon form positive definite. The physical Hilbert space is its completion,
\begin{align}\label{eq:scalar-physical-space}
    {\cal H}_{\rm phys}^{\rm sc}
    :=&
    \overline{
        V^{\rm sc}/V_{\rm g}^{\rm sc}
    } \,,
    \\[3pt]
    \Pi_{\rm phys}^{\rm sc}
    \simeq&\Pi_{p,0} \,.
\end{align}
Thus, a physical scalar state is an equivalence class of modes in $V^{\rm sc}$ modulo the finite-dimensional gauge sector.

The successive layers of the covariant representation may be displayed schematically as
\begin{align}
    V_{\rm tot}^{\rm sc}/V^{\rm sc}
    \;\longrightarrow\;
    \underbrace{
        V^{\rm sc}/V_{\rm g}^{\rm sc}
    }_{\Pi_{p,0}}
    \;\longrightarrow\;
    \underbrace{
        V_{\rm g}^{\rm sc}
    }_{\pi_{p,0}} \,,
\end{align}
with Hilbert completion understood for the physical quotient. The arrows encode the non-trivial extensions under the $\mathrm{dS}_4$ group action, rather than maps between these subquotients. The physical UIR therefore occurs as the central Gupta--Bleuler quotient, rather than as an invariant direct summand of $V_{\rm tot}^{\rm sc}$.

%%%%%%%%%%%%%%%%%%%%%%%%%%%%%%%%%%%%%%%%%%%%%%%%%%%%%
%%%%%%%%%%%%%%%%%%%%%%%%%%%%%%%%%%%%%%%%%%%%%%%%%%%%%
\subsubsection{Spherical Realization on $\mathbb{S}^3$}

The spherical realization of $\Pi_{p,0}$ is constructed by removing a finite-dimensional sector of low harmonics and equipping the resulting quotient with an invariant inner product. Let $Y_{Llm}$ denote the hyperspherical harmonics on $\mathbb S^3$, and define
\begin{align}
    V_{p-1}
    :=&
    \operatorname{span}
    \left\{
        Y_{Llm}\;\colon\;0\leqslant L\leqslant p-1
    \right\} \,,
    \\[3pt]
    H_{p-1}
    :=&
    V_{p-1}^{\perp_{L^2}} \,.
\end{align}
The orthogonal decomposition
\begin{align}
    L^2_{\mathbb C}(\mathbb S^3)
    =
    H_{p-1}\oplus V_{p-1}
\end{align}
identifies $H_{p-1}$ with a space of representatives
of the quotient,
\begin{align}
    H_{p-1}
    \simeq
    L^2_{\mathbb C}(\mathbb S^3)/V_{p-1} \,.
\end{align}
The group action on these representatives is understood as the induced quotient action; the $L^2$-orthogonal decomposition does not imply a decomposition into invariant summands.

Since the degree-$L$ harmonic space has dimension $(L+1)^2$, the sector removed has dimension
\begin{align}
    \dim V_{p-1}
    =
    \sum_{L=0}^{p-1}(L+1)^2
    =
    \frac{p(p+1)(2p+1)}{6}
    =
    \dim V_{\rm g}^{\rm sc} \,.
\end{align}
This agrees with the dimension of the spacetime gauge sector and suggests the correspondence between the two reductions, although dimension counting alone does not establish their representation-theoretic identification.

The physical Hilbert structure is defined by a non-local invariant form, rather than the ordinary $L^2$ product. For smooth representatives $f_1,f_2\in H_{p-1}$, it is
\begin{align}\label{eq:scalar-spherical-inner-product}
    \langle f_1,f_2\rangle_{p-1}
    ={}&
    \frac{(-1)^{p+1}}{4\pi^2(2p-1)!}
    \iint_{\mathbb S^3\times\mathbb S^3}
    f_1^\ast(u_1)\,f^{}_2(u_2)
    \nonumber\\[3pt] 
    &\qquad\times
    |u_1-u_2|^{2(p-1)}
    \log|u_1-u_2|^{-2}
    \,\mathrm{d}u_1\,\mathrm{d}u_2 \,.
\end{align}
Here, $|u_1-u_2|$ is the Euclidean chordal distance on the unit sphere. The form is positive definite on these representatives and invariant under the induced quotient action.

To express the same construction on unrestricted smooth functions, let $P_p$ be the $L^2$-orthogonal projection onto $H_{p-1}$ and define
\begin{align}
    \langle[f_1],[f_2]\rangle_{p-1}
    :=
    \langle P_pf_1,P_pf_2\rangle_{p-1} \,.
\end{align}
Equivalently, if $K_p$ denotes the integral operator in Eq. \eqref{eq:scalar-spherical-inner-product}, the quotient form is represented by $P_pK_pP_p$. This projection ensures independence of the representatives without requiring the unprojected logarithmic kernel to annihilate the low-harmonic sector.

Completing the smooth representative space in the norm $\|f\|_{p-1}^2:=\langle f,f\rangle_{p-1}$ gives
\begin{align}
    \widehat H_{p-1}
    :=
    \overline{
        H_{p-1}\cap C^\infty(\mathbb S^3)
    }^{\,\|\cdot\|_{p-1}} \,,
\end{align}
which carries the UIR $\Pi_{p,0}$. This completion is taken in the invariant norm and need not coincide with the ordinary $L^2$ completion. Below, when $H_{p-1}$ is used as the representation carrier, this invariant Hilbert completion is understood.

%%%%%%%%%%%%%%%%%%%%%%%%%%%%%%%%%%%%%%%%%%%%%%%%%%%%%
%%%%%%%%%%%%%%%%%%%%%%%%%%%%%%%%%%%%%%%%%%%%%%%%%%%%%
\subsubsection{From the Spacetime Realization to the Conformal Boundary}

The conformal-boundary behavior of the scalar modes connects the spacetime and spherical constructions. In global conformal coordinates, the two boundaries are
\begin{align}
    \mathscr I^\pm
    :=
    \left\{\rho=\pm\frac{\pi}{2}\right\}
    \simeq\mathbb S^3 \,,
\end{align}
and boundary data are extracted from the asymptotic behavior as $\rho\to\pm\pi/2$, with the rescaling and normalization appropriate to the scalar modes \cite{Bajalan2026}.

The boundary construction must respect the gauge equivalence encoded by Eq. \eqref{GBScalar}. The finite-dimensional bulk gauge sector $V_{\rm g}^{\rm sc}$ gives rise to the low-harmonic sector $V_{p-1}$, so that bulk fields differing by a gauge mode determine boundary data differing by an element of $V_{p-1}$. Passing to equivalence classes therefore yields the correspondence
\begin{align}
    V_{\rm g}^{\rm sc}
    &\longrightarrow V_{p-1} \,,
    \\[3pt]
    V^{\rm sc}/V_{\rm g}^{\rm sc}
    &\longrightarrow
    L^2_{\mathbb C}(\mathbb S^3)/V_{p-1}
    \simeq H_{p-1} \,,
\end{align}
initially on the corresponding mode spaces, with invariant Hilbert completion understood. The choice of representatives in $H_{p-1}$ removes the low-harmonic ambiguity without changing the boundary equivalence classes.

This correspondence identifies the physical quotients; it does not identify the full bulk Gupta--Bleuler filtration with the $L^2$-orthogonal decomposition $H_{p-1}\oplus V_{p-1}$. In particular, the complementary bulk quotient $V_{\rm tot}^{\rm sc}/V^{\rm sc}$ is not needed to define the physical boundary carrier.

With the appropriate normalization, the induced boundary map transports the Klein--Gordon inner product on the physical bulk quotient to the invariant form $\langle\cdot,\cdot\rangle_{p-1}$. Its extension to the Hilbert completions thus identifies the physical conformal-boundary realization with the spherical realization of $\Pi_{p,0}$. The resulting unitary equivalence is expressed by the intertwiner described next.

%%%%%%%%%%%%%%%%%%%%%%%%%%%%%%%%%%%%%%%%%%%%%%%%%%%%%
%%%%%%%%%%%%%%%%%%%%%%%%%%%%%%%%%%%%%%%%%%%%%%%%%%%%%
\subsubsection{Unitary Equivalence of the Spherical and Spacetime Realizations}

The two constructions are related by the explicit unitary equivalence established in Ref. \cite{Bajalan2026}. We denote by $\mathcal F_p$ the intertwiner in the bulk-to-boundary direction:
\begin{align}
    \mathcal F_p \colon\;
    \left(
        {\cal H}_{\rm phys}^{\rm sc},\;
        \Pi_{p,0}^{\mathrm{dS}},\;
        \langle\cdot,\cdot\rangle_{\rm KG}
    \right)
    \;\xrightarrow{\;\simeq\;}\;
    \left(
        \widehat H_{p-1},\;
        \Pi_{p,0}^{\mathbb S^3},\;
        \langle\cdot,\cdot\rangle_{p-1}
    \right) \,.
\end{align}
Here, $\widehat H_{p-1}$ is the invariant Hilbert completion of the spherical carrier, and $\langle\cdot,\cdot\rangle_{\rm KG}$ denotes the positive-definite form induced on the physical spacetime quotient and extended to its completion. The map $\mathcal F_p$ is surjective and intertwines the group actions,
\begin{align}
    \mathcal F_p\,
    \Pi_{p,0}^{\mathrm{dS}}(\texttt{g})
    =
    \Pi_{p,0}^{\mathbb S^3}(\texttt{g})\,
    \mathcal F_p \,,
    \qquad
    \texttt{g}\in\mathrm{SO}_0(1,4) \,,
\end{align}
while preserving the invariant inner products,
\begin{align}
    \left\langle
        \mathcal F_p\Phi_1,\mathcal F_p\Phi_2
    \right\rangle_{p-1}
    =
    \langle\Phi_1,\Phi_2\rangle_{\rm KG} \,,
    \qquad
    \Phi_1,\Phi_2\in{\cal H}_{\rm phys}^{\rm sc} \,.
\end{align}
Thus, $\mathcal F_p$ transports the physical spacetime group action and Hilbert structure to the spherical realization, while its inverse reconstructs physical bulk states from boundary data.

Together with the boundary analysis, this result relates the finite-dimensional sectors removed in the two constructions and identifies their physical carriers: \begin{align}
    V_{\rm g}^{\rm sc}
    &\longleftrightarrow V_{p-1} \,,
    \\[3pt]
    {\cal H}_{\rm phys}^{\rm sc}
    &\xrightarrow[\;\simeq\;]{\mathcal F_p}
    \widehat H_{p-1} \,.
\end{align}
The first correspondence concerns the gauge sectors; the second is the unitary equivalence after reduction and completion. These are the structures that the non-scalar program must recover, with the additional task of constructing a boundary realization from the interior of $\mathbb B$.

%%%%%%%%%%%%%%%%%%%%%%%%%%%%%%%%%%%%%%%%%%%%%%%%%%%%%
%%%%%%%%%%%%%%%%%%%%%%%%%%%%%%%%%%%%%%%%%%%%%%%%%%%%%
\subsection{From the Bounded Realization to the Boundary: the Non-Scalar Bosonic Problem} \label{subsec:nonscalar-compact-boundary}

The scalar intertwiner $\mathcal F_p$ identifies the physical spacetime realization with an independently constructed spherical realization. For the non-scalar bosonic representations $\Pi^\pm_{p,q}$, $1\leqslant q\leqslant p$, the analogous program requires an additional step: the bounded realization is carried by sections over the interior of $\mathbb B$, so its boundary carrier must first be identified. The two bulk--boundary problems are
\begin{align}
    \mathbb B
    \;\overset{?\mathrm{hol}}{\longleftrightarrow}\;
    \partial\mathbb B \,,
    \qquad
    \mathrm{dS}_4
    \;\overset{?\mathrm{hol}}{\longleftrightarrow}\;
    \mathscr I^\pm \,.
\end{align}
Each construction must produce a boundary realization of the chosen UIR on $\mathbb S^3$, together with its invariant Hilbert structure. Their explicit identification then completes the correspondence proposed in Eq. \eqref{eq:program}.

We begin with the established bounded realization
\begin{align}
    \left(
        {\mathcal H}_{p,q}^{\pm}(\mathbb B),\;
        \Pi_{p,q}^{\mathbb B,\pm},\;
        \langle\cdot,\cdot\rangle_{p,q}
    \right) \,,
\end{align}
whose inner product is given in Eq. \eqref{eq:compact-inner-product}. In the chosen bundle realization, its sections take values in $V_{\tau_{p,q}^{\pm}}$. The radial parametrization 
\begin{align}
    z=ru \,,
    \qquad
    0\leqslant r<1 \,,
    \qquad
    u\in\mathbb S^3
\end{align}
provides a geometric approach to $\partial\mathbb B$ as $r\to1^{-}$. Extracting the appropriate boundary data may require rescaling and fibre projection, together with constraints or quotient relations. These operations must be compatible with the induced group action and the invariant boundary norm.

The first task is therefore to construct a unitary intertwiner
\begin{align}
    {\mathcal B}_{p,q}^{\pm}\colon\;
    \left(
        {\mathcal H}_{p,q}^{\pm}(\mathbb B),\;
        \Pi_{p,q}^{\mathbb B,\pm},\;
        \langle\cdot,\cdot\rangle_{p,q}
    \right)
    \;\xrightarrow{\;\simeq\;}\;
    \left(
        {\mathcal H}_{p,q}^{\partial,\pm}(\mathbb S^3),\;
        \Pi_{p,q}^{\partial,\pm},\;
        \langle\cdot,\cdot\rangle_{\partial;p,q}
    \right) \,.
\end{align}
As for $\mathcal F_p$, the direction is from bulk to boundary. The map must be surjective and satisfy
\begin{align}\label{eq:nonscalar-boundary-intertwining}
    {\mathcal B}_{p,q}^{\pm}\,
    \Pi_{p,q}^{\mathbb B,\pm}(\texttt{g})
    =
    \Pi_{p,q}^{\partial,\pm}(\texttt{g})\,
    {\mathcal B}_{p,q}^{\pm} \,,
    \qquad
    \texttt{g}\in\mathrm{SO}_0(1,4) \,,
\end{align}
together with
\begin{align}\label{eq:nonscalar-boundary-isometry}
    \left\langle
        {\mathcal B}_{p,q}^{\pm}f_1,
        {\mathcal B}_{p,q}^{\pm}f_2
    \right\rangle_{\partial;p,q}
    =
    \langle f_1,f_2\rangle_{p,q} \,.
\end{align}
Its inverse reconstructs the bounded realization from boundary data; the Szeg\"o-type constructions discussed above provide a starting point for this inverse direction.

The scalar prototype suggests that the invariant boundary inner product may be non-local. Suppose that boundary representatives take values in a finite-dimensional space $E_{p,q}^{\pm}$. This fibre must be determined as part of the construction and need not coincide with the bulk fibre $V_{\tau_{p,q}^{\pm}}$. A candidate boundary form can then be expressed through a distributional kernel
\begin{align}
    {\mathcal K}_{p,q}^{\pm}
    \in
    {\mathcal D}^{\prime}
    \left(
        \mathbb S^3\times\mathbb S^3,\,
        \operatorname{End}(E_{p,q}^{\pm})
    \right)
\end{align}
as
\begin{align}\label{eq:nonscalar-kernel-product}
    \langle F_1,F_2\rangle_{\partial;p,q}
    =
    \iint_{\mathbb S^3\times\mathbb S^3}
    \left\langle
        F_1(u),
        {\mathcal K}_{p,q}^{\pm}(u,v)F_2(v)
    \right\rangle_{E_{p,q}^{\pm}}
    \,\mathrm{d}u\,\mathrm{d}v \,.
\end{align}
For a non-trivial boundary bundle, this expression is understood in local trivializations, with the kernel acting between the corresponding fibres. The form must descend through any required quotient and become positive definite on the resulting carrier, whose completion defines ${\mathcal H}_{p,q}^{\partial,\pm}(\mathbb S^3)$. The projected logarithmic kernel of the scalar construction provides a structural model for this procedure, rather than a $q\to0$ specialization of the non-scalar kernel.

Covariance provides one route to determining this Hilbert structure. If the action on boundary representatives is
\begin{align}
    \bigl(
        \Pi_{p,q}^{\partial,\pm}(\texttt{g})F
    \bigr)(u)
    =
    J_{p,q}^{\pm}(\texttt{g},u)\,
    F(\texttt{g}^{-1}\cdot u) \,,
\end{align}
invariance of Eq. \eqref{eq:nonscalar-kernel-product} imposes a bi-covariance condition on ${\mathcal K}_{p,q}^{\pm}$ involving both the fibre cocycle $J_{p,q}^{\pm}$ and the transformation of the boundary measure. When the physical action requires projection or passage to a quotient, this condition must hold on the resulting carrier. The boundary action, constraints, and invariant form must therefore be determined together.

A complementary approach uses the harmonic decomposition of the bounded realization. Let
\begin{align}
    \left\{
        \Phi_{\alpha}^{\pm;p,q}
    \right\}_{\alpha\in{\mathcal A}_{p,q}}
\end{align}
be an orthonormal basis of ${\mathcal H}_{p,q}^{\pm}(\mathbb B)$ adapted to the maximal compact subgroup. One seeks boundary extraction operators $T_r^{\pm;p,q}$ such that
\begin{align}
    \Psi_{\alpha}^{\pm;p,q}
    :=
    \lim_{r\to1^-}
    T_r^{\pm;p,q}
    \left[
        \Phi_{\alpha}^{\pm;p,q}(r\,\cdot)
    \right]
\end{align}
exists in an appropriate distributional or Sobolev sense. These operators may incorporate rescaling and fibre projection. If they act diagonally in the harmonic decomposition, their normalization factors must still be compatible with the full $\mathrm{SO}_0(1,4)$ action.

The resulting boundary classes should form a complete orthonormal system,
\begin{align}
    \left\langle
        \Psi_{\alpha}^{\pm;p,q},
        \Psi_{\beta}^{\pm;p,q}
    \right\rangle_{\partial;p,q}
    =
    \delta_{\alpha\beta} \,.
\end{align}
The boundary map would then be given by
\begin{align}
    {\mathcal B}_{p,q}^{\pm}f
    =
    \sum_{\alpha\in{\mathcal A}_{p,q}}
    \Psi_{\alpha}^{\pm;p,q}\,
    \left\langle
        \Phi_{\alpha}^{\pm;p,q},f
    \right\rangle_{p,q} \,,
\end{align}
with convergence in the boundary Hilbert norm. Orthonormality and completeness yield a surjective isometry; the intertwining identity \eqref{eq:nonscalar-boundary-intertwining} additionally requires the two systems to transform compatibly under the group.

In a compatible integral realization of the bulk inner product, this expansion suggests the formal transform kernel
\begin{align}
    \mathbb K_{p,q}^{\pm}(u,z)
    :=
    \sum_{\alpha\in{\mathcal A}_{p,q}}
    \Psi_{\alpha}^{\pm;p,q}(u)\,
    \Phi_{\alpha}^{\pm;p,q}(z)^\dagger \,.
\end{align}
Its use incorporates the bulk measure and fibre metric entering $\langle\cdot,\cdot\rangle_{p,q}$, and requires a suitable convergence statement. The transform kernel $\mathbb K_{p,q}^{\pm}(u,z)$ and the boundary inner-product kernel ${\mathcal K}_{p,q}^{\pm}(u,v)$ thus serve distinct roles.

On the spacetime side, the direct analogue of $\mathcal F_p$ must extract asymptotic data consistently with the Gupta--Bleuler reductions described above. In particular, gauge-equivalent bulk representatives must determine the same physical boundary class. For the graviton, this requirement applies to both the ordinary gauge and scalar zero-mode reductions.

Fix one conformal boundary, either $\mathscr I^+$ or $\mathscr I^-$, and one irreducible component $\Pi^\pm_{p,q}$. The superscript $\pm$ on the representation and its carrier labels the UIR and is independent of the choice of conformal boundary. The required unitary map is
\begin{align}
    {\mathcal A}_{p,q}^{\pm}\colon\;
    \left(
        {\cal H}_{p,q}^{{\rm phys},\pm},\;
        \Pi_{p,q}^{\mathrm{dS},\pm},\;
        \langle\cdot,\cdot\rangle_{\rm phys}
    \right)
    \;\xrightarrow{\;\simeq\;}\;
    \left(
        {\mathcal H}_{p,q}^{\mathscr I,\pm}(\mathbb S^3),\;
        \Pi_{p,q}^{\mathscr I,\pm},\;
        \langle\cdot,\cdot\rangle_{\mathscr I;p,q}
    \right) \,.
\end{align}
It must satisfy
\begin{align}
    {\mathcal A}_{p,q}^{\pm}\,
    \Pi_{p,q}^{\mathrm{dS},\pm}(\texttt{g})
    &=
    \Pi_{p,q}^{\mathscr I,\pm}(\texttt{g})\,
    {\mathcal A}_{p,q}^{\pm} \,,
    \\[3pt]
    \left\langle
        {\mathcal A}_{p,q}^{\pm}\Phi_1,
        {\mathcal A}_{p,q}^{\pm}\Phi_2
    \right\rangle_{\mathscr I;p,q}
    &=
    \langle\Phi_1,\Phi_2\rangle_{\rm phys} \,.
\end{align}
These are precisely the counterparts of the intertwining and isometry properties of $\mathcal F_p$. The additional work lies in identifying the non-scalar boundary data and implementing the appropriate gauge reduction.

Once the two boundary constructions carry the same UIR, their abstract unitary equivalence follows. The middle problem in Eq. \eqref{eq:program} is to construct its explicit implementation:
\begin{align}
    {\mathcal U}_{p,q}^{\pm}\colon\;
    {\mathcal H}_{p,q}^{\partial,\pm}(\mathbb S^3)
    \;\xrightarrow{\;\simeq\;}\;
    {\mathcal H}_{p,q}^{\mathscr I,\pm}(\mathbb S^3) \,.
\end{align}
This map must intertwine the boundary actions and
preserve their invariant inner products,
\begin{align}
    {\mathcal U}_{p,q}^{\pm}\,
    \Pi_{p,q}^{\partial,\pm}(\texttt{g})
    &=
    \Pi_{p,q}^{\mathscr I,\pm}(\texttt{g})\,
    {\mathcal U}_{p,q}^{\pm} \,,
    \\[3pt]
    \left\langle
        {\mathcal U}_{p,q}^{\pm}F_1,
        {\mathcal U}_{p,q}^{\pm}F_2
    \right\rangle_{\mathscr I;p,q}
    &=
    \langle F_1,F_2\rangle_{\partial;p,q} \,.
\end{align}
For each fixed irreducible component, such a unitary intertwiner is unique up to phase.

If an explicit unitary bulk intertwiner
\begin{align}
    W_{p,q}^{\pm}\colon\;
    {\mathcal H}_{p,q}^{\pm}(\mathbb B)
    \longrightarrow
    {\cal H}_{p,q}^{{\rm phys},\pm}
\end{align}
is available, compatibility can be expressed by
\begin{align}
    {\mathcal U}_{p,q}^{\pm}\,
    {\mathcal B}_{p,q}^{\pm}
    =
    {\mathcal A}_{p,q}^{\pm}\,
    W_{p,q}^{\pm} \,,
\end{align}
or, equivalently,
\begin{align}
    {\mathcal U}_{p,q}^{\pm}
    =
    {\mathcal A}_{p,q}^{\pm}\,
    W_{p,q}^{\pm}\,
    \bigl({\mathcal B}_{p,q}^{\pm}\bigr)^{-1} \,.
\end{align}
The two routes then produce the same physical boundary data: one through the bounded boundary realization, the other through the physical spacetime realization. Constructing these maps with compatible normalizations would join the two bulk--boundary correspondences and complete the chain proposed in Eq. \eqref{eq:program}.

%%%%%%%%%%%%%%%%%%%%%%%%%%%%%%%%%%%%%%%%%%%%%%%%%%%%%
%%%%%%%%%%%%%%%%%%%%%%%%%%%%%%%%%%%%%%%%%%%%%%%%%%%%%
%%%%%%%%%%%%%%%%%%%%%%%%%%%%%%%%%%%%%%%%%%%%%%%%%%%%%
\section{Conclusion}

We have formulated a twofold holographic program for the non-scalar bosonic discrete series of $G=\mathrm{SO}_0(1,4)$, guided by the established scalar correspondence. The central task is to construct normalized unitary boundary maps for the bounded and spacetime realizations and explicitly intertwine the resulting boundary representations. The vector, graviton, and partially massless graviton provide concrete tests of this program, with their gauge structures determining the reductions through which the physical boundary carriers must emerge.

The framework distinguishes the symmetric-space carrier $G/\mathrm{SO}(4)$, its associated coadjoint-orbit fibrations, and Lorentzian $\mathrm{dS}_4$. Their different geometric roles are brought together at the level of the physical representations and their invariant Hilbert structures. Establishing the proposed intertwiners would reveal which features of the scalar correspondence extend across the non-scalar bosonic discrete series and which depend on spin and gauge depth. Beyond the one-particle theory, these results could provide a foundation for investigating a holographic formulation of quantum field theory in $\mathrm{dS}_4$, including its strictly massless, partially massless, and gravitational sectors.

%%%%%%%%%%%%%%%%%%%%%%%%%%%%%%%%%%%%%%%%%%%%%%%%%%%%%
%%%%%%%%%%%%%%%%%%%%%%%%%%%%%%%%%%%%%%%%%%%%%%%%%%%%%
\section*{Acknowledgements}

J.-P.G. and H.P. thank the University of Valladolid for its hospitality during the 36th International Colloquium on Group Theoretical Methods in Physics (Group36), and Mariano A. del Olmo for his warm hospitality and for his efforts in organizing the conference.

%%%%%%%%%%%%%%%%%%%%%%%%%%%%%%%%%%%%%%%%%%%%%%%%%%%%%
%%%%%%%%%%%%%%%%%%%%%%%%%%%%%%%%%%%%%%%%%%%%%%%%%%%%%
\paragraph{Funding information}

M.B. and H.P. acknowledge support from the National Science Fund, Ministry of Education and Science of Bulgaria, under contract KP-06-N92/2.

%%%%%%%%%%%%%%%%%%%%%%%%%%%%%%%%%%%%%%%%%%%%%%%%%%%%%
%%%%%%%%%%%%%%%%%%%%%%%%%%%%%%%%%%%%%%%%%%%%%%%%%%%%%
\bibliography{references}

\end{document}